\documentclass[12pt]{article}
\usepackage[dvips]{graphicx}
\usepackage{amssymb}
\usepackage{amsmath}
\def\dd{{\rm d}}\def\ee{{\rm e}}\def\ii{{\rm i}}

\def\beq{\begin{equation}}\def\eeq{\end{equation}}
\def\bea{\begin{eqnarray}}\def\eea{\end{eqnarray}}

\begin{document}
 
\title{``Classical" model of discrete QFT:\\ Klein-Gordon and electromagnetic fields}
 
\author{Roman Sverdlov, Institute of Mathematical Sciences,
\\IV IISER Mohali, Knowledge city, Sector 81, SAS Nagar, Manauli} 

\date{Feburary, 2013}
\maketitle
 
\begin{abstract}
\noindent  The purpose of this paper is to propose a ``classical" model of \emph{quantum} fields which is local. Yet it admittedly violates relativity as we know it and, instead, it fits within a bimetric model with one metric corresponding to speed of light and another metric to superlumianl signals whose speed is still finite albeit very large. The key obstacle to such model is the notion of functional in the context of QFT which is inherently non-local. The goal of this paper is to stop viewing functionals as fundamental and instead model their emergence from the deeper processes that are based on functions over $\mathbb{R}^4$ alone. The latter are claimed to be local in the above bimetric sense. 
\end{abstract}

\subsection*{Terminology and conventions}

$\; \; \; \; \;$ {\bf Lower case latin indexes} denote a lattice point. For example, $A^{\mu}_k$ is the value of the field $A^{\mu}$ at a lattice point $k$.

{\bf Upper case latin indexes} denote choice of metric (O=''ordinary'', L= ''lower ordinary'', U=''upper ordinary'', S=''superluminal''). For example, $A_L^{\mu}$, $A_U^{\mu}$ and $A_S^{\mu}$ are metric transformations of each other, as described in Section 4.

{\bf Important modification} Greek indexes are transforming according to superluminal metric $g_{S; \mu \nu}$ (instead of ordinary $g_{O; \mu \nu}$) which would imply superluminal, but finite, speed of light. Anything related to our ordinary speed of light is emergent rather than fundamental. 

{\bf LPL particles}, which stands for ''lattice pointlike particles'', will be the term we will use to denote lattice points. That is due to the non-trivial dynamical properties we attribute to lattice points per Sections 3 and 6. \emph{They are not to be confused with physical particles}

{\bf Physical particles} represent wavelike processes occuring on LPL-particle background (LPL particle to water molecule is the same as physical particle is to water wave). Their description is given in Section 2.

\subsection*{1. Introduction}

It is commonly understood that quantum non-locality takes place during measurement. It should be pointed out, however, that the issues of relativity and locality are separate. Our intuition does not demand that the speed of signal be $3 \times 10^8 m/s$, but it does demand that it be finite. In fact, there were some proposals of bimetric framework designed to accommodate superluminal signals which still move with finite speed (see \cite{Bimetric1}, \cite{Bimetric2} and \cite{Bimetric3}). In this paper we propose that bimetric theory emerges from more fundamental, single metric one; where the "fundamental metric" is superluminal. We argue that our universe is a crystal-like structure that lives in an underlying continuum space, with the speed $c$ being the outcome of the specific crystal structure, similar to the speed of sound, while the fundamental speed is much higher than $c$; we will call these two speeds ``ordinary'' and ``superluminal'', and denote them by $c_{\rm O}$ and $c_{\rm S}$, respectively. Let's assume for simplicity that the size $L$ of the crystal is finite; likewise, the necessary time for measurement to occur, $\delta \tau$, is finite as well. If we assume that $c_{\rm S} \gg L/ \delta \tau$, it would imply that all of the necessary communication can take place within the duration of the relevant measurements, leading to the desired entanglement.

 Admittedly, this violates relativity as we know it. But we can simply claim that  $c_{\rm O}$-based relativity is merely the result of the physics of the crystal and does not hold on a more fundamental level.  For example, electric and magnetic fields are fundamentally different but they simply happened to co-exist within the crystal, thus leading to false appearance of relativity (since they are analogous to sound rather than light, neither of them exist outside of crystal). On the other hand, $c_{\rm S}$-based relativity holds fundamentally, but it is broken by the reference frame of the crystal -- similarly to how $c_{\rm O}$-based relativity is broken by the reference frame of the air in which sound propagates. Thus, the combination of $c_{\rm S}$ and $c_{\rm O}$ will in fact determine a preferred frame as the common central line of two light cones, but such preferred frame would be specific to the crystal that is needed for emergence of $c_{\rm O}$ and will coincide with the velocity of the crystal. Outside the crystal, the $c_{\rm S}$-based relativity without any preferred frame will hold.

What we have said so far is that measurement, as such, does not imply non-locality. However, there are more fundamental factors that do, which persist even in measurement-free scenario. In particular, the wave function is defined on a configuration space rather than ordinary space. In order for configuration space to be mathematically well defined, we need a strict notion of simultaneity as opposed to an approximate one. In other words, the speed of superluminal signals has to be strictly infinite. This problem persists even in the framework of quantum field theory. If we are to attempt to define probability amplitude as something evolving in time (as opposed to in- and out- states at $t = \pm \infty$), we would be required to select a preferred foliation into hypersurfaces and then define probability amplitudes as functionals over the infinite-dimensional space of $(\phi, A^{\mu})$ on each hypersurface. Arguably, we can preserve relativity by selecting all possible hypersurfaces at the expense of allowing redundancies; yet the notion of functional over each hypersurface will be distinctly non-local. The key role of configuration space in creating conceptual problem has been widely acknowledged by leading scientists of the time (some of their quotes can be found at \cite{PhysicsForums}).

We can attempt to solve this problem by claiming that a functional is merely an emergent outcome of more fundamental physics based on ordinary functions in $\mathbb{R}^4$. In this context, we can re-introduce the finite speed of superluminal signals while function-based physics will continue to be well defined; the only price to pay is that the correspondence to functional-based physics will now be approximate rather than exact, while we claim that the error is too small to be detected. However, the obvious obstacle on our way is the fact that the set of functionals is far larger than the set of functions, which makes it seemingly impossible to model the former in terms of the latter. We propose to answer this question by claiming that the domain of our functional does not consist of all conceivable functions, but rather it consists of a specific subset of functions that has emerged within our crystal. The finiteness of that set makes it technically smaller than $\mathbb{R}^3$ and therefore embeddable within the latter. On the other hand, if the set of functionals produced by the crystal is bounded, then it is conceivable that the finite set of functions will be packed closely enough to look continuous. 

We propose that the selected set of functions emerges in the following way. Each LPL particle can emit or absorb superluminal signals of a specific type, identified with the values of a numeric parameter $I_q$, dictated by the value of a ``tuning parameters'' $q$ and $q'$ attached to any given point ($I_q$ plays a similar role to frequency in radio waves, although in our case the frequencies of all waves are the same). We will further impose the constraint that the values of $q$ are restricted to integer values between $1$ and $M$. Now, if the number of LPL particles in our crystal is $N \gg M$, there would be a very large number of LPL particles (approximately $N/M$) that share any given value of $q$; the latter we will refer to as $q$-th sublattice. Now, if we assign to each LPL particle internal parameters $\phi$ and $A^{\mu}$, then there will be a one-to-one correspondence between each sublattice and a specific $(\phi, A^{\mu})$, obtained by pointwise collecting the readings of the latter from each LPL particle of said sublattice. Thus, we will obtain a selection of finite set of functions that we will utilize in the definition of our functional. 

The logical next step is to identify the functional over $(\phi, A^{\mu})$ with a complex valued function over the set of sublattices. The latter is still non-local. But it can be made local by assigning $\psi$ to each individual LPL particle on a sublattice and then utilizing superluminal signals to make sure that the LPL particles of any given sublattice somehow copy the values of $\psi$ from each other. Due to the finiteness of the speed of superluminal signals the values of $\psi$ at those LPL particles cannot be exactly equal, but they can still be approximately equal. Up to said order of approximation, we can identify the common value of $\psi$ across \emph{the LPL particles} on a given sublattice with the assignment of $\psi$ to that sublattice \emph{as a whole}; the latter will then be identified with the value of the functional $\psi (\phi, A^{\mu})$ for the specific $(\phi, A^{\mu})$ associated with that sublattice. By repeating this argument for all sublattices we will, indeed, produce the discretized version of a functional we are looking for. 

One way to understand the above proposal is by an analogy with a hologram. The pictures drawn on a hologram do not change in time. What changes in time is our choice of the picture we are looking at. Similarly, the values of $(\phi, A^{\mu})$ drawn on each sublattice do not change with time; but our choice of the sublattice we want to look at does. In order to reproduce the classical physics, we have to ``look'' at one specific sublattice and ``not look'' at any other one; in other words, the values of $\psi$ at the elements of a given sublattice should be $1$ while the values of $\psi$ at all other LPL particles should be $0$; and then as time progresses the $1$ across one sublattice will become $0$ while the $0$ across another sublattice will become $1$ which will cause the perceived change in picture; but at the same time neither $\phi$ nor $A^{\mu}$ changes on any given LPL particle; it is the value of $\psi$, namely $0$ or $1$, that does. 

In order to ``quantize'' the above, we have to relax the zero-or-one assumption by allowing $\psi$ to have any other complex value. At the same time, we retain the assumption that $\psi$ should be the same across any given sublattice (which will be enforced by means of superluminal signals). The common value of the latter will be a identified with the ``probability amplitude'' that we in fact chose to ``look'' at the sublattice in question. This set of probability amplitudes corresponds to the functional we are trying to reproduce. The synchronization across sublattice is due to superluminal signals that pass across the crystal within negligible time period. The reason that the given time period is negligible is that the time evolution of $\psi$ that occur within crystal are progressing a lot slower, which is why the information we get from the delayed superluminal signal is still reliable. 

In order to keep the hologram continuous-looking, we have to introduce upper and lower bounds for the values of the fields. Thus, by having a large enough number of bounded field distributions we would statistically expect some portion of them to approximate any given field distribution we imagine with said bounds, up to some tolerance $\epsilon_{\phi}$. We argue that this does not lead to contradictions with quantum field theory. In the case of a scalar field, the analogy with the harmonic oscillator tells us that very large values of $\phi$ represent the tail of the probability distribution, and cutting off said tail has negligible effect. The situation becomes more critical in gauge theories where, due to symmetries, we expect the probability amplitude to be constant among an unbounded set of gauge equivalent distributions. Nevertheless, the Fadeev-Popov procedure will guarantee that the outcome is independent of either the shape or the presence of the boundaries of the region of integration. Thus we will continue to agree with QFT even in this context. 

It has to be emphasized, however, that the dynamics that governs the motion of the particles of the crystal is quite arbitrary. Most of the equations I present are simply examples of one kind of dynamics that would accomplish the above purposes, but the same goals can be accomplished by other kinds of dynamics as well. This paper is simply a counter-example to the claim that it is impossible to describe quantum field theory as an emergent outcome of some local classical processes. There are, in principle, other equally good counter examples. Nevertheless, if one is to replace wave equations of propagation of signals and instead describe the absorption as immediate outcome of emission, one would obtain cellular automaton whose algorithm of operation is less arbitrary. In future research it might be possible to explore what happens on this more qualitative level. But for the purposes of this paper we have decided to include the detailed differential equations that are both arbitrary and unnecessary simply in order to demonstrate that \emph{if one insists} on this, this can be done. 

\subsection*{2. Correspondence between functionals and Fock space}

So far we have shown how to classically model the appearance of functionals. However, the typical object of interest in QFT calculation is the probability amplitude of a state in a Fock space. The analogy with the harmonic oscillator tells us that there should be a correspondence between these two notions and, therefore, we \emph{should} be able to define Fock space. Let us demonstrate how we do that. 

For notational convenience, we will assign numbers to LPL particles. Thus, if the lattice as a whole has $N$ LPL particles, each LPL particle is identified with $k \in \{1, \cdots, N\}$. We will denote the value of the tuning parameter $q$ of a LPL particle $k$ by $q_k$. Similarly, we will denote the values of $(\phi, A^{\mu})$ and $\psi$ attached to a LPL particle $k$ by $(\phi_k, A_k^{\mu})$ and $\psi_k$, respectively. On the other hand, the restriction of $(\phi, A^{\mu})$ to the $q$-th sublattice will be denoted by $(\phi_{(q)}, A^{\mu}_{(q)})$, where brackets around the index are used to indicate that the index refers to a sublattice rather than an individual LPL particle. 

Let us now attempt to model a typical quantum field theory state. In order for the situation to be manageable, we will focus on states that can be written as \emph{finite} sums of products of creation operators corresponding to different momenta $p$, acting on the vacuum. The set of included momenta is to be appropriately chosen so as to satisfy the boundary conditions but, if we consider real fields $\phi$, for each value of $p$ it makes sense to include $-p$ in the set. Therefore, we propose our generic state to be of the form
\beq
\vert \psi \rangle = \sum_{\{c\},\{d\}} K_{c_1 \cdots c_n d_1 \cdots d_n} (a_{p_1}^{\dagger})^{c_1} (a_{- p_1}^{\dagger})^{d_1} \cdots (a_{p_n}^{\dagger})^{c_n} (a_{-p_n}^{\dagger})^{d_n} \vert 0 \rangle\;. \label{GenericState}
\eeq
This state corresponds to a wave function 
\beq
\psi(\phi) = \sum_{\{c\},\{d\}} \Bigg[ K_{c_1 \cdots c_n d_1 \cdots d_n} \prod_a \lambda_{c_a d_a} \Bigg(\int \dd^3 x \,\phi (x) \cos (\vec{p} \cdot \vec{x}),  \int \dd^3 x\, \phi(x) \sin(\vec{p} \cdot \vec{x}) \Bigg) \Bigg], \label{FockScalar}
\eeq
where 
\bea
& &\lambda_{cd}(x,y) = \Bigg[\prod_{j=1}^n \bigg( \frac{m^2+p_j^2}{4} \Bigg)^{\!1/4} (x + \ii y) + \frac{1}{\sqrt{2}(m^2+p^2)^{1/4}} \bigg(\frac{\dd}{\dd x} + \ii \frac{\dd}{\dd y} \bigg) \Bigg)^{\!c_j} \Bigg] \times\\
& &\times\ \Bigg[\prod_{j=1}^n \Bigg( \bigg( \frac{m^2+p_j^2}{4} \bigg)^{\!1/4} (x - \ii y) + \frac{1}{\sqrt{2}(m^2+p^2)^{1/4}} \bigg(\frac{\dd}{\dd x} - \ii\, \frac{\dd}{\dd y} \bigg) \Bigg)^{\!d_j} \Bigg] \ee^{-\frac{\sqrt{m^2+p^2}}{2} (x^2 + y^2)} \;. \nonumber
\eea
We now rewrite the above state as a ``wave function" on a Hilbert space, $\psi (\phi)$. First, we replace $\psi(\phi)$ with $\psi(q)$, where $q$ is a ``hologram" describing $\phi$ by $q_k = q \Rightarrow \phi_k = \phi (\vec{x}_k)$, for all $k$. We then replace $\psi(q)$ with $\psi_k$, using $\psi_k = \psi(q_k)$ for all $k$. Thus, the state given in Eq.\ \ref{GenericState}, by definition, corresponds to the situation where every single LPL particle $k$ simultaneously satisfies
\beq
\psi_k = \sum_{\{c\},\{d\}} \Bigg[ K_{c_1 \cdots c_n d_1 \cdots d_n} \prod_a \lambda_{c_a d_a} \Bigg(\sum_{l \in A_{q_k}} \phi_l \cos (\vec{p} \cdot \vec{x}_l),  \sum_{l \in A_{q_k}} \phi_l \sin (\vec{p} \cdot \vec{x}_l) \Bigg) \Bigg], \label{FockScalar1}
\eeq
In the case of photons, one can simply rewrite Eq.\ \ref{FockScalar1} replacing $\phi$ with $A^{\mu}$. It is important to note that, while $\partial_0$ and $\partial_k$ are fundamentally different, $A^0$ and $A^k$ are a lot more similar. At the kinematical level, the latter does not refer to either evolution in space or in time, which is why the causal nature of $t$ becomes irrelevant, 
\beq
\psi_k = \sum_{\{c\},\{d\}} \Bigg[ K_{c_1 \cdots c_n d_1 \cdots d_n} \prod_a \prod_{j=1}^2 \lambda_{c_a d_a} \Bigg(\sum_{l \in A_{q_k}} A_l^{\mu} e_{j \mu} (p) \cos (\vec{p} \cdot \vec{x}_l),  \sum_{l \in A_{q_k}} A_l^{\mu} e_{j \mu} (p) \sin (\vec{p} \cdot \vec{x}_l) \Bigg) \Bigg],
\eeq
where $e_{1 \mu} (p)$ and $e_{2 \mu} (p)$ are the bases for the description of the polarization of a photon with momentum $p^{\mu}$; thus,
\beq
p^{\mu} e_{1 \mu}(p) = p^{\mu} e_{2 \mu}(p) = e_1^{\mu}(p) e_{2 \mu}(p) = 0\;.
\eeq

\subsection*{3. Formation of the lattice and relativity-related issues}

So far we have convinced ourselves that the mathematical information about Fock space can, indeed, be read off from a ``crystal" living in ordinary space, at least at the kinematical level. This means that we are now motivated enough to describe in more detail how the crystal is formed. The latter will be needed in order to substantiate our analogy between $c_{\rm O}$ and the speed of sound that we have discussed earlier.  As we stated earlier, we introduce two fundamental speeds, $c_{\rm O}$ and $c_{\rm S}$ (``O" stands for ``ordinary" and ``S" stands for ``superluminal") satisfying
\beq
3 \times 10^8\;\hbox{m/s}
\approx c_{\rm O} \ll \frac{L}{\delta \tau} \ll c_{\rm S} < \infty\;,
\eeq
where $L$ is the size of a box our universe is enclosed in and $\delta \tau$ is a very small time interval; both $L$ and $\delta \tau$ are finite. The inequality $c_{\rm O} < c_{\rm S}$ implies the violation of relativity; on the other hand, the inequality $c_{\rm S} < \infty$ implies locality. The ``non-local" nature of emergent functionals is made possible by the fact that $L/ \delta \tau \ll c_{\rm S}$, which makes $c_{\rm S}$ ``appear" infinite even if it isn't.

As we said earlier, our crystal is embedded in a larger continuum space. This will allow us to claim that only the former has finite size while the latter is infinite. Thus, the superluminal signal passes the crystal producing the needed changes, and then flies away to infinity. In order to explain the finite size of the crystal, we will propose that it is growing; thus its size parameters are direct consequences of the time it had to grow so far. By the time it ``grows" more, it will become ``too large" for superluminal signals to pass it within a very small time, and then quantum field theory will break down. But we assume that this will only happen in the far future. 

We will set up our proposed growth model in the following way. LPL particles will be viewed as possessing a ``charge" as they interact through an electromagnetic-like potential $V^{\mu}$ (which has nothing to do with the ``actual" electromagnetic field $A^{\mu}$; it simply has a similar mathematical structure) that obeys
\beq
\partial_{\mu} (\partial^{\mu} V^{\nu} - \partial^{\nu} V^{\mu}) = \sum_{k=1}^N \int_{\gamma_k} \delta^4 (x - \gamma_k (\tau))\, \dd \tau\;, \label{maxwelllike1}
\eeq
where there are $N$ LPL particles (numbered $k=1, \cdots, N$) and $x^{\mu} = \gamma_k^{\mu} (\tau)$ is the trajectory of LPL particle number $k$ in spacetime. The worldline of an LPL particle is determined by that vector potential; but it is quite different from what one would expect in electrodynamics. First of all, the LPL particles are assumed to have the ``same charge" (namely $+1$); secondly, the ``same charge" interaction has an attractive effect at large distances and a repulsive effect at close distances, which would lead to the formation of a lattice with a ``preferred" distance scale. This can be accomplished by imposing the following dynamical equation:
\beq
\frac{\partial^2 x^{\mu}}{\partial \tau^2} = a\, g_{\rm S}^{\mu\nu} \,
\ee^{\beta\, \dd(V^{\nu}V_{\nu} - W^2)^2/\dd\tau} \partial_{\perp v; \mu} (V^{\nu} V_{\nu} - W^2)^2 \; ; \quad v^{\mu} = \frac{\dd x^{\mu}}{\dd\tau} \;,\label{motion}
\eeq
where, for any given $v^{\mu}$, the ``parallel derivative" $\partial_{\parallel v; \mu}$ and the ``orthogonal derivative" $\partial_{\perp v}^{\mu}$ are given by 
\beq
\partial_{\parallel v; \mu} f = \frac{g_{S; \mu\rho} v^{\rho} v^{\sigma} \partial_{\sigma} f}{g_{S; \mu \nu} v^{\mu} v^{\nu}} \; ; \quad \partial_{\perp v; \mu} = \partial_{\mu} f - \frac{g_{S; \mu \rho} v^{\rho} v^{\sigma} \partial_{\sigma} f}{g_{S; \eta \chi} v^{\eta} v^{\chi}}\;. \label{Orth1}
\eeq
The above definitions of ``parallel" and ``orthogonal" derivatives can be motivated by the observation that, in the $(t,x,y,z)$-convention, 
\beq
v = (v, 0, 0, 0) \Rightarrow \big[\partial_{\parallel v} f = (\partial_0 f, 0, 0, 0, 0) \; ; \; \partial_{\perp v} f = (0, \partial_1 f, \partial_2 f, \partial_3 f) \big] \label{Orth2}\;.
\eeq 
The effect of the above potential is to attract any given LPL particle $j$ to the location 
\beq
g_{S; \mu \nu} V^{\mu} (x_j)  V^{\nu} (x_j) \approx W^2\;.
\eeq
and then not allowing it to leave it due to the irreversibility resulting from the exponential factor. If we impose initial conditions 
\beq
x^0 = 0\ \Rightarrow\ V^{\mu} (x) = 0\;, \label{Initial}
\eeq
then $g_{S; \mu \nu} V^{\mu} (x_j)  V^{\nu} (x_j) \approx W^2$ will be satisfied on a ``preferred" distance scale. At the same time, when the LPL particles are far away from each other, the attractive force is not terribly strong.  Thus, they are moving around at their initial velocity and whenever they happen to pass by each other by accident, they get ``stuck" to each other. 

Now, if the number of LPL particles is finite, then the infinite size of the universe implies arbitrarily large distance between any two LPL particles, which would imply zero probability of the formation of the lattice structure. On the other hand, if we assume that the number of LPL particles is infinite, then any given LPL particle will be subject to an infinitely large superluminal influence from far away. Take, for example, Eq.\ (\ref{maxwelllike1}). This equation implies that the contribution towards $V^{\mu}$ from the distant LPL particles behaves like $1/r$. This means that the contribution from the LPL particles on a surface of radius $r$ is of the order of $r$. Therefore, the integral over all $r$ will produce infinity.  In order to avoid this problem, we will have to introduce a damping parameter. The first impulse is to simply introduce a $\partial V^{\mu} / \partial t$ term, in a ``preferred" time $t$. But, for aesthetic reasons, we would rather maintain $c_{\rm S}$-based relativity (even though it would still violate the $c_{\rm O}$-based one). Thus, instead of $\partial V^{\mu} / \partial t$, we will use $g_{\rho \sigma} V^{\rho} V^{\mu} \partial_{\mu} V^{\sigma}$, and Eq.\ (\ref{maxwelllike1}) becomes 
\beq
\partial_{\mu} (\partial^{\mu} V^{\nu} - \partial^{\nu} V^{\mu}) + \epsilon_d \, g_{\rho \sigma} V^{\rho} V^{\mu}\, \partial_{\mu} V^{\sigma} = \sum_{k=1}^N \int_{\gamma_k} \delta^4 (x - \gamma_k (\tau))\, \dd \tau\;, \label{maxwelllike2}
\eeq
where "d" in $\epsilon_d$ stands for "dissipation". The above depends on the assumption that $V^{\mu}$ is timelike, and has positive time component. This is a consequence of the initial conditions (\ref{Initial}), together with the fact that $v^{\mu} = dx^{\mu} / d \tau$ is timelike with positive time component. We will introduce similar damping components towards any other wave equations we will be dealing with in the next sections:
\beq
g_{\rm S}^{\alpha \beta} \partial_{\alpha} \partial_{\beta} \mu \mapsto g_{\rm S}^{\alpha \beta} \partial_{\alpha} \partial_{\beta} \mu - \epsilon_d \, V^{\alpha} \partial_{\alpha} \mu\;.
\eeq
This will allow us to say that we have infinitely many LPL particles in the universe, but the influence from ``far away" LPL particles is arbitrarily small. This will further allow us to assume that the average number of LPL particles per unit volume on a spacelike hypersurface is finite rather than infinitely small. However, we can assume that the average distance between LPL particless (which is likewise finite) is much larger than the one required for equilibrium, per Eq.\ (\ref{maxwelllike2}). Thus, they still have to ``run into each other" in order to form a lattice. But this time the probability of this happening is non-zero. This further implies that, if we ``wait long enough", the lattice will form with absolute certainty. In fact, infinitely many lattice structures will be forming; we claim that we are living in one of them. 

Now, we claim that the entire universe that we live in is just one of these several lattice structures. The coefficient $\epsilon_d$ in Eq.\ (\ref{maxwelllike2}) is so small that the size of the lattice structure we are living in is not large enough for its effect to be felt. Thus, we will be using Eq.\ (\ref{maxwelllike1}) as a close approximation throughout the rest of the paper. At the same time, however, the nearest LPL particle \emph{outside} our lattice will be expected to be ``far enough" for the $\epsilon_d$-term to be significant. This apparent statistical contradiction can be accommodated by an assumption that the average \emph{finite} distance between any \emph{two} given LPL particles is ``much larger" than the size of the entire lattice we are living in; and that is despite the fact that the latter includes billions of LPL particles! More precisely, if the number of LPL particles in the lattice we are living in is $N(t)$, and if the average density is $\rho_{\rm in}$ for LPL particles inside the lattice and $\rho_{\rm out}$ outside, then 
\beq
{\rm Our \; Time \; Period} \Rightarrow N(t)\, \rho_{\rm out} \ll \rho_{\rm in}\;.
\eeq
The discrepancy between $\rho_{\rm out}$ and $\rho_{\rm in}$ is due to their very different origins. The density $\rho_{\rm out}$ is entirely based on the original distribution of LPL particles and is independent of their dynamics; while $\rho_{\rm in}$ is an ``equilibrium density" determined entirely from the \emph{dynamical} Eq.\ (\ref{maxwelllike2}). Now, since the above drastic discrepancy is still finite, the lattice is still guaranteed to form if we wait long enough. Once the lattice has formed, it is being held together per either Eq.\ (\ref{maxwelllike1}) or (\ref{maxwelllike2}) (which closely approximate each other on this scale). At the same time, nothing ``holds" any of the LPL particles outside the lattice. Thus, the nearest LPL particle \emph{outside} the lattice is separated by an expected distance of the order of $1/\rho_{\rm out}^3$, which is several magnitudes larger than the size of the lattice. This will allow us to assume that $\epsilon_d$ is ``large enough" for the total effect of \emph{all} LPL particles outside the lattice to be negligible and, at the same time, $\epsilon_d$ is ``small enough" for Eq.\ (\ref{maxwelllike1}) to be a very close approximation to Eq.\ (\ref{maxwelllike2}) \emph{inside} the lattice. 

It should be noticed that the spacing between LPL particles might not be constant, due to the effects any given LPL particle experiences from ``far away" LPL particles \emph{in the same lattice}. In particular, closer to the edges of the lattice the spacing might end up being different than it is towards the center. Now, as was pointed out in Ref.\ \cite{Zee}, one can expect the ultraviolet cutoff to be inversely proportional to the spacing of LPL particles. Thus, we might expect the ultraviolet cutoff to change as we move across the lattice; this would ultimately imply variations of the renormalized mass, charge, and so forth, assuming that the bare parameters are the same. However, we can claim that the part of the universe that is accessible to our observations is ``very small" compared to the size of the lattice; thus, the variation of the ultraviolet cutoff is negligible within the region accessible to our observations.  

This, however, assumes that the spacing between LPL particles approximates a continuous function. This, too, can be questioned. It is possible that small-scale interactions would lead to non-trivial distance variations even on few-LPL scales. Nevertheless, one can still expect that the pattern of discontinuities would form some kind of repeated structure. This structure might average to something continuous. In other words, we might have three ``very small" space scales $\delta_1$, $\delta_2$ and $\delta_3$ which satisfy $\delta_1 \ll \delta_2 \ll \delta_3$. We can have discontinuities on the scale of $\delta_1$ which would average to something continuous on the scale of $\delta_2$. A function that is ``continuous" on the scale $\delta_2$ will, in fact, vary on the scale of $\delta_3$; yet, $\delta_2$ is too small for variation to occur, which makes the function nearly-constant on that scale. We can then assume that $\delta_3 \ll L$ and \emph{despite that}, $\delta_2$ is the scale of the observable universe. This will allow a variation of the ultraviolet cutoff throughout the lattice (whose scale is $L$) while at the same time that variation would be ``too far away" for us to see.  

In this framework it is possible to argue that the theory of relativity \emph{is} satisfied, after all. However, the version of relativity that is satisfied is $c_{\rm S}$-based rather than $c_{\rm O}$-based. On the other hand, $c_{\rm O}$ is a parameter that strictly applies to a lattice that has already been formed. Thus, the lattice identifies the ``preferred frame" in much the same way as the water identifies a ``preferred frame" for the propagation of water waves; and, just like the water waves move slower than the speed of light $c_{\rm O}$, photons move much slower than the speed $c_{\rm S}$. Furthermore, since different lattices can form at the same time, they will be moving relative to each other. Thus, each lattice will carry its own ``preferred frame"; just like the geological processes of the Earth are based on Earth's preferred frame while similar processes in the Moon are based on the frame of the Moon. This illustrates that in reality neither lattice is a true preferred frame. Instead, we have a strict relativity, based on $c_{\rm S}$. 

It should be pointed out, however, that the belief in $c_{\rm S}$-based relativity is logically independent of any of the experiments we have performed that taught us relativity. After all, it is perfectly conceivable that \emph{despite} $c_{\rm S}$-based relativity, the processes we see within a given lattice demonstrate the violation of relativity (just like the processes we see in the ocean do). The reason relativity is respected is that we have ``cleverly designed" each lattice in such a way that it is (this has been done in Sec.\ 5 and 6). For example, it could have been possible to have electricity without magnetism, which would have demonstrated violation of relativity; but we ``cleverly" introduced magnetism in order to ``hide" this. Our ability to ``design" a lattice in such a way that it respects $c_{\rm O}$-based relativity is independent of the validity of the $c_{\rm S}$-based one, as evident from the difference between the values of $c_{\rm O}$ and $c_{\rm S}$. Our insistence on the validity of $c_{\rm S}$-based relativity is only aesthetic and is not backed up by any experimental evidence.  

\subsection*{4. Bimetric relativity on a continuum}

So far we have said that in the empty space we have $c_S$-based relativity; the LPL-particles, on the other hand, form a medium in which $c_O$-based relativity holds. This means that $c_S$ based processes are on a continuum while $c_O$ based ones are discrete. However, if we take a continuum limit of the latter, we will arrive at a bimetric continuum theory. Nevertheless, in light of our underlying knowledge of how $c_O$ arises, we would like $c_S$-based metric to be more fundamental than $c_O$ based one. In other words, throughout this paper we will raise and lower indexes by means of $g_S^{\mu \nu}$. This in particular means that we are using $g_S$ in raising the indexes of $g_O$,
\beq g_O^{\mu \nu} = g_S^{\mu \rho} g_{O; \rho \sigma} g_S^{\sigma \nu} \eeq
But in order to reproduce $g_O$-based relativity, we would like to be able to raise $g_O$ indexes by means of $g_O$ itself. In order to do that, we have to think of the "lower index" version of $g_O$ and "upper index" one as two separate tensor fields (just like $F_{\mu \nu}$ and $G_{\mu \nu}$ are completely separate) and simply postulate that the product of these tensors (defined in $g_S$-based framework) gives $\delta^{\mu}_{\nu}$. Thus, we will denote these two separate tensors by $g_L$ and $g_U$, where "L" stands for "lower" and "U" stands for "upper". Thus, 
\beq g_U^{\mu \rho} g_{L; \rho \nu} = \delta^{\mu}_{\nu} \label{gugl}\eeq
Now, since we are "taking seriously" the fact that $g_L$ and $g_U$ are two separate tensors, we should be able to raize the indexes of $g_L$ by means of $g_S$ to obtain $g_L^{\mu \nu}$ while it is still $g_L$ rather than $g_U$. Similarly, we should be able to lower the indexes of $g_U$ to get $g_{U; \mu \nu}$ while it is still $g_U$ rather than $g_L$: 
\beq g_L^{\mu \nu} = g_S^{\mu \rho} g_{L; \rho \sigma} g_S^{\sigma \nu} \; , \; g_{U; \mu \nu} = g_{S; \mu \rho} g_U^{\rho \sigma} g_{S; \sigma \nu} \eeq
Now in ordinary relativity we expect to raise and lower vector indexes by means of $g_O$. In our case, we view $g_S$ as fundamental which means that vector indexes are raised by means of the latter, 
\beq A^{\mu} = g_S^{\mu \nu} A_{\nu} \eeq
In order to reproduce raising/lowering indexes by means of $g_O$, we have to again introduce two different vector fields, $A_U$ and $A_L$, related through
\beq A_U^{\mu} = g_U^{\mu \nu} A_{L; \nu} \; , \; A_{L; \mu} = g_{L; \mu \nu} A_U^{\nu} \eeq
Once again, each $A_U$ and $A_L$ are vector fields on their own right, which means that one can use $g_S$ to raise and lower indexes of these fields while leaving $u$ and $l$ unchanged: 
\beq A_U^{\mu} = g_S^{\mu \nu} A_{U \nu} \; , \; A_{U \mu} = g_{S; \mu \nu} A_U^{\nu} \label{No-index-to-index1} \eeq
\beq A_L^{\mu} = g_S^{\mu \nu} A_{L\nu} \; , \; A_{L \mu} = g_{S; \mu \nu} A_L^{\nu} \label{No-index-to-index2} \eeq
We can now use the above in order to relate $A_U$ and $A_L$ while freely choosing between upper and lower indexes as we wish: 
\beq A_U^{\mu} = g_U^{\mu \rho} g_{S; \rho \nu} A_L^{\nu}  \; , \; A_{U; \mu} = g_{S; \mu \rho} g_U^{\rho \nu} A_{L; \nu} \eeq
\beq A_L^{\mu} = g_S^{\mu \rho} g_{L; \rho \nu} A_U^{\nu}  \; , \; A_{L; \mu} = g_{L; \mu \rho} g_S^{\rho \nu} A_{U; \nu} \eeq
\beq A_{U; \mu} = g_{S; \mu \rho} g_U^{\rho \sigma} g_{S; \sigma \nu} A_L^{\nu} \; , \; A_L^{\mu}=g_S^{\mu \rho} g_{L; \rho \sigma} g_S^{\sigma \nu} A_{U; \nu} \eeq
\beq A_U^{\mu} = g_U^{\mu \nu} A_{L ; \nu} \; , \; A_{L; \mu} = g_{L ; \mu \nu} A_U^{\nu} \eeq
Similarly, the tensor fields can be shown to obey 
\beq T_U^{\mu \nu} = g_U^{\mu \alpha} g_U^{\nu \beta} g_{S; \alpha \rho} g_{S; \beta \sigma} T_L^{\rho \sigma} \; , \; T_{U; \mu \nu} = g_{S; \mu \alpha} g_{S; \nu \beta} g_U^{\alpha \rho} g_U^{\beta \sigma} T_{L; \rho \sigma} \eeq
\beq T_L^{\mu \nu} = g_S^{\mu \alpha} g_S^{\nu \beta} g_{L; \alpha \rho} g_{L; \beta \sigma} T_U^{\rho \sigma} \; , \; T_{L; \mu \nu} = g_{L; \mu \alpha} g_{L; \nu \beta} g_S^{\alpha \rho} g_S^{\beta \sigma} T_{U; \rho \sigma} \eeq
\beq T_{U; \mu \nu} = g_{S; \mu \chi} g_{S; \nu \eta} g_U^{\chi \alpha} g_U^{\eta \beta} g_{S; \alpha \rho} g_{S; \beta \sigma} T_L^{\rho \sigma} \eeq
\beq T_L^{\mu \nu} = g_S^{\mu \alpha} g_S^{\nu \beta} g_{L; \alpha \chi} g_{L; \beta \eta} g_S^{\chi \rho}  g_S^{\eta \sigma} T_{U; \rho \sigma} \eeq
\beq T_{L; \mu \nu} = g_{L; \mu \rho} g_{L; \nu \sigma} T_U^{\rho \sigma} \; , \; T_U^{\mu \nu} = g_U^{\mu \rho} g_U^{\nu \sigma} T_{L; \rho \sigma} \eeq
The other ingredient is the definition of covariant derivatives. In the above discussion we were viewing $A_L$ and $A_U$ as two separate vector fields. Having more than one vector field is, of course, something we already used to. On the other hand, we have only one coordinate system which means only one definition of a derivative, $\partial_{\mu}$. But, for our convenience, we will identify $\partial_{L; \mu}$ with $\partial_{\mu}$ and then use our vector prescription to define $\partial_L^{\mu}$, $\partial_U^{\mu}$ and $\partial_{U; \mu}$: 
\beq \partial_{L ; \mu}  = \partial_{\mu} \; , \; \partial^{\mu} = \partial_L^{\mu} = g_S^{\mu \nu} \partial_{\nu}   \; , \; \partial_{U; \mu} = g_{S; \mu \rho} g_U^{\rho \nu} \partial_{\nu} \;  , \;  \partial_U^{\mu} = g_U^{\mu \nu} \partial_{ \nu} \eeq
It is easy to check that the relations we had for $A_L$ and $A_S$ continue to hold for $\partial_L$ and $\partial_S$. Let us now reproduce scalar electrodynamic Lagrangian in this new notation. The Lagrangian for the interaction is 
\beq
{\cal L} = m^2 \phi^* \phi + {\cal D}_U^{\mu} \phi^* {\cal D}_{L; \mu} \phi + F_U^{\mu \nu} F_{L; \mu \nu} \; , \; {\cal D}_U^{\mu} \phi = \partial_U^{\mu} \phi + \ii eA_U^{\mu} \phi \; ; \; {\cal D}_U^{\mu} \phi^* = \partial_U^{\mu} \phi^* - \ii eA_U^{\mu} \phi^*\;.
\label{LagrCont}\eeq
Upon some simple algebra, the above expression becomes
\beq
{\cal L}= F_U^{\mu \nu} F_{L; \mu \nu} + \partial_U^{\mu} \phi^* \partial_{L; \mu} \phi + e^2 A_U^{\mu} A_{L; \mu} \phi^* \phi + \ii eA_U^{\mu} (\phi \partial_{L; \mu} \phi^* - \phi^* \partial_{L; \mu} \phi) + m^2 \phi^2\;.
\eeq
By remembering the expression for the current,
\beq
j_U^{\mu} = \ii e (\phi \partial_U^{\mu} \phi^* - \phi^* \partial_U^{\mu} \phi)\;,
\eeq
it is easy to see that the $ie$ term is the interaction between the ``current" and the photon. In fact, the only ``surprising" term is a four-vertex $A^{\mu} A_{\mu} \phi^* \phi$. Upon further thought it can be realized that this term is not that surprising either: while such four-vertex is ``forbidden" for spin $1/2$ fermions (since the latter have dimension $3/2$), it is ``allowed" for spin $0$ boson (which has dimension $1$). 

Now, the existence of two different metrics implies preferred frame. For example, in 1+1 dimensions one can single out the $t$-axis to be the common center of both light cones. On the other hand, the $g_O$-based relativity (which includes both $g_L$ and $g_U$) only emerges in the interior of the crystal discussed in the previous section, whereas $g_S$ based relativity holds regardless. This should give us a clue that the "preferred frame" needed for bimetric theory arises from the velocity of LPL-particles. The $c_S$-based relativity is maintained because LPL particles are \emph{not} stationary (which would have dictated preferred frame if they were) but instead they dynamically form quasi-stationary structures as described in the previous section. An LPL particle number $k$ has a velocity $v_k^{\mu}$ which, in a continuum limit, turns into $v^{\mu} (x)$. We can now define $g_L$ and $g_U$ according to
\beq g_{L; \mu \nu} = v_{\mu} v_{\nu} \bigg(\frac{c_S}{c_O} - 1 \bigg) + g_{S; \mu \nu} \; , \;  g_U^{\mu \nu} = v^{\mu} v^{\nu} \bigg(\frac{c_O}{c_S}-1 \bigg) +g_S^{\mu \nu} \label{Toyglgu}\eeq 
It is easy to verify that the above definitions of $g_L$ and $g_U$ obey Eq. \ref{gugl}, which is the only requirement that they have to satisfy. However, in the context of future work where we will define gravity, the above equation is tentative. It is conceivable that we might want $g_S$ to remain flat while $g_L$ and $g_U$ to be curved, in which case Eq \ref{Toyglgu} needs to be abandoned. But of course it is also possible to explore the possibility of adding curvature to $g_S$ as well in which case Eq \ref{Toyglgu} might still be satisfied. For the purposes of this paper we are only dealing with flat space. So, as a toy model. we will assume that Eq \ref{Toyglgu} holds. 

If we now identify $A^{\mu}$ with $A_U^{\mu}$ and use Eq \ref{Toyglgu}, we will be able to re-express everything in terms of $g_S$-covariant quantities:
\beq A_U^{\mu} = A^{\mu} \; , \; A_{L; \mu} =  A^{\nu} v_{\mu} v_{\nu} \bigg(\frac{c_S}{c_O} - 1 \bigg) + A_{\mu} \eeq
\beq \partial_{L \mu} = \partial_{\mu} \; , \; \partial_U^{\mu} = \bigg(\frac{c_O}{c_S}-1 \bigg)  v^{\mu} v^{\nu} \partial_{\nu} + \partial^{\mu} \eeq
where both $A^{\mu}$ and $\partial_{\mu}$ are being raised and lowered by means of $g_S$:
\beq A_{\mu} = g_{S; \mu \nu} A^{\nu} \; , \; A^{\mu} = g_S^{\mu \nu} A_{\nu} \; , \; \partial_{\mu} = g_{S; \mu \nu} \partial^{\nu} \; , \; \partial^{\mu} = g_S^{\mu \nu} \partial_{\nu} \eeq
By substituting this into Eq \ref{LagrCont}, we obtain
\beq
{\cal L}= \bigg(\bigg(\frac{c_O}{c_S}-1 \bigg)  (v^{\mu} v^{\rho} \partial_{\rho} A^{\nu} - v^{\nu} v^{\rho} \partial_{\rho} A^{\mu} ) + \partial^{\mu} A^{\nu} - \partial^{\nu} A^{\mu} \bigg) \times \nonumber \eeq
\beq \times \bigg(  \bigg(\frac{c_S}{c_O} - 1 \bigg) (v_{\nu} v_{\rho} \partial_{\mu} A^{\rho} - v_{\mu} v_{\rho} \partial_{\nu} A^{\rho}) + \partial_{\mu} A_{\nu} - \partial_{\nu} A_{\mu} \bigg) + \nonumber \eeq
\beq + \bigg(\frac{c_O}{c_S}-1 \bigg)  v^{\mu} v^{\nu} \partial_{\nu} \phi^* \partial_{\mu} \phi + \partial^{\mu}  \phi^* \partial_{\mu} \phi +  \label{csAlone} \eeq
\beq + e^2 A^{\mu} A^{\nu} \phi^* \phi v_{\mu} v_{\nu} \bigg(\frac{c_S}{c_O} - 1 \bigg) +  e^2 A^{\mu} A_{\mu} \phi^* \phi + \nonumber \eeq
\beq + \ii eA^{\mu} (\phi \partial_{\mu} \phi^* - \phi^* \partial_{\mu} \phi) + m^2 \phi^2\;. \nonumber
\eeq
It is important to point out that in the above expression we have gotten rid of all of the $U$-s and $L$-s, while all of the Lorentz contractions are $c_S$-based. In other words, the expression has only one relativistic covariance, the one of $c_S$. That covariance is broken by the preferred frame set up by $v^{\mu}$. This is particularly the case due to the fact that we treat $A^{\mu}$ and $\phi$ as the only variables while treating $v^{\mu}$ as afore-given. On the other hand, the idea that $v^{\mu}$ obeys the dynamics of previous section allows us to continue to view it as physical field, although on a different standing. In this respect $c_S$-based relativity is still preserved. 

For the purposes of the rest of the paper, we will substitute Eq \ref{csAlone} into $c_S$-only framework; this will allow us to forget about $g_L$ and $g_U$ altogether. 

Just like in standard theory of relativity there are rules of producing Lorentz covariant expressions, in our case we have rules as well:

1. Write down a Lagrangian in such a way that all indexes contract in two different ways at the same 

a) For each upper $\mu$ there is lower $\mu$, and visa versa (it should be understood, however, that this type of contraction is $c_S$-based)

b) For each $U$-based quantity with $\mu$-index there is $L$-based quantity with same index, and visa versa

2. Substitute Equations \ref{No-index-to-index1} and \ref{No-index-to-index2}, thus get rid of $L$-s and $U$-s

3. Re-interpret the result as $c_S$-based \emph{alone} and view the peculiar structure as mere coincident. 

\subsection*{5. Discretized Lagrangian density} 

So far we have introduced a continuum Lagrangian density. However, since LPL-particles form a discrete structure, the latter needs to be discretized. In light of the fact that the structure is not cubic, we can not simply replace derivatives with differences. Instead, we need to define derivatives statistically.  We introduce a \emph{Lagrangian generator} ${\cal K} (x^{\mu}, y^{\mu}, z^{\mu}; {\rm fields})$ depending on three spacetime points, then define a \emph{Lagrangian density} ${\cal L} (\vec{z}, t)$ as 
\beq
{\cal L}(z^{\mu};{\rm fields}) = \int_{S(z, v)} \dd^d x\, \dd^d y\,
{\cal K}(x^{\mu} - v^{\mu} (x) \delta \tau, y^{\mu} - v^{\mu} (y) \delta \tau,
z^{\mu} ; {\rm fields} )\;, \label{LagrDensity}
\eeq
where $S (z^{\mu}, v^{\mu})$ is a surface passing through $z^{\mu}$ which is perpendicular to $v^{\mu}$ at every point. We will assume that the value of ${\cal K}$ is very small \emph{unless} all three points it acts upon are very close to each other. Thus, the above integral is approximately equal to similar integral over a small patch of the above surface, which can be assumed to be unique. 

It turns out that there is a ``mechanical" way of translating a Lagrangian involving a given combination of derivatives into a Lagrangian generator, as we will now see. Let us define operator $\Gamma_{\mu \alpha}$ (where $\mu$ is Lorentz index and $\alpha$ is a very large real number), that will replace the ordinary derivatives $\partial_{\mu}$ and a function $I_\alpha$, by: 
\beq
(\Gamma_{\mu \alpha} f) (x^{\mu}, z^{\mu}) = \frac{\alpha^{d/2}}{(2 \pi)^{d/2} }\, \, (f(z) -  f(x)) \bigg(\frac{v^{\mu}}{\delta \tau} + \frac{\alpha}{\pi} (z^{\mu} - x^{\mu} - v^{\mu} v^{\nu} (z_{\nu} - x_{\nu})) \bigg) \times
\eeq
\beq \times \exp \bigg( \frac{\alpha}{2} \, ( z^{\nu}- x^{\nu} - v^{\nu} v^{\rho} (z_{\rho} - x_{\rho}))( z_{\nu}- x_{\nu} - v_{\nu} v^{\sigma} (z_{\sigma} - x_{\sigma}))\bigg) \eeq
\beq
I_{\alpha} (x^{\mu}, z^{\mu}) = \Big(\frac{\alpha}{2 \pi}\Big)^{d/2}  \exp \bigg( \frac{\alpha}{2} \, ( z^{\nu}- x^{\nu} - v^{\nu} v^{\rho} (z_{\rho} - x_{\rho}))( z_{\nu}- x_{\nu} - v_{\nu} v^{\sigma} (z_{\sigma} - x_{\sigma}))\bigg) 
\eeq
where we have $e^{+ \alpha/2 \cdots}$ instead of $e^{- \alpha/2 \cdots}$ due to the extra minus sign which comes from squaring spacelike vector inside the exponent in $(+,-,-,-)$ convention; $f$ stands for any function on spacetime (for example, $f = \phi$ or $f = A^\mu$). Now, as we recall, we are living in a crystal in which the velocities of the particles are approximately parallel to each other. We will set the time axis to coincide with the common direction of velocity vectors; thus,
\beq v^{\mu} \approx \delta^{\mu}_0 \eeq
Furthermore, we will assume that $f$ changes much slower than the time it takes for superluminal signal to cross the crystal. This implies that we can treat superluminal signal as infinitely fast and still obtain close approximation to the results we are seeking. These considerations imply
\beq
(\Gamma_{0 \alpha} f) (x^{\mu}, z^{\mu}) = \frac{\alpha^{d/2}}{(2 \pi)^{d/2} \delta \tau}\, \ee^{-\alpha\, \vert\vec{z}-\vec{x}\vert^2/2}\, (f(z^{\mu}) -  f(x^{\mu}))
\eeq
\beq
 (\Gamma_{p\alpha} f) (\vec{x}, \vec{z}) = \frac{2 \alpha^{(d+2)/2}}{(2 \pi)^{(d+2)/2}}\, \ee^{-\alpha\, \vert\vec{z}-\vec{x}\vert^2/2}\, (z^p - x^p) (f(z^{\mu}) - f(x^{\mu}))
\eeq
\beq
I_{\alpha} (x^{\mu}, z^{\mu}) = \Big(\frac{\alpha}{2 \pi}\Big)^{d/2} \ee^{-\alpha\, \vert\vec{z}-\vec{x}\vert^2/2}\;,
\eeq
Let us now try a Lagrangian generator of the form 
\beq
{\cal K} (x^{\mu}, y^{\mu}, z^{\mu}; \phi) = [(\Gamma_{0 \alpha} \phi)(x^{\mu}, z^{\mu})]  [(\Gamma_{p\alpha} \phi) (y^{\mu}, z^{\mu})]\;, \label{toy1}
\eeq
where we have violated rotational symmetry by selecting a ``preferred" axis $p$. By substituting the expressions for the $\Gamma$s, we obtain
\beq
{\cal K} (x^{\mu}, y^{\mu}, z^{\mu}; \phi) = \Big[\frac{\alpha^{d/2}}{(2 \pi)^{d/2} \delta \tau}\, \ee^{-\alpha\, \vert \vec{z} - \vec{x} \vert^2/2}\, (\phi (z^{\mu}) -  \phi (x^{\mu})) \Big] \times \nonumber
\eeq
\beq
\times \Big[ \frac{2 \alpha^{(d+2)/2}}{(2 \pi)^{(d+2)/2}}\, \ee^{-\alpha\, \vert \vec{z} - \vec{y} \vert^2/2}\, (z^k - y^k) (\phi (z^{\mu}) - \phi (y^{\mu})) \Big]\,. \label{toy3}
\eeq
Now, by inspecting Eq.\ \ref{LagrDensity}, we see that the ingredients inside the Lagrangian generator are ``taken" at a different time. In particular, $x^0 = y^0 = z^0 - \delta \tau$. Therefore,  Eqs.\ \ref{toy3} produces
\beq
{\cal L} = \int \dd^d x\, \dd^d y \Big\{ \Big[\frac{ \alpha^{d/2}}{(2 \pi)^{d/2} \delta \tau} \,\ee^{-\frac{\alpha}{2} \vert \vec{z} - \vec{x} \vert^2} \Big( \phi (\vec{z}, t) - \Big( \phi (\vec{x}, t) -  \delta \tau \,\frac{\partial \phi}{\partial t} \Big\vert_{\vec{x}} \Big) \Big) \Big] \times
\eeq
\beq
\times \Big[ \frac{2 \alpha^{(d+2)/2}}{(2 \pi)^{(d+2)/2}} \,\ee^{- \frac{\alpha}{2} \vert \vec{y} - \vec{z} \vert^2} (y^p-z^p) \Big(\phi (\vec{z}, t) - \Big(\phi (\vec{y}, t) - \delta \tau \,\frac{\partial \phi}{\partial t} \Big\vert_{\vec{y}} \Big)\Big)\Big]\Big\}\;,
\eeq
Now, if $\alpha \gg 1$, then we can assume that the only values of $\vec{x}$ and $\vec{y}$ that make a non-negligible contribution to the integral are the ones that are very close to $\vec{z}$. Therefore, if we assume that $\phi$ is well behaved, we can assume that $\phi$ is linear in the region where its contribution is non-negligible:
\beq
\phi (\vec{x}, t) \approx \phi (\vec{z}, t) + (\vec{x} - \vec{z}) \cdot \vec{\nabla} \phi \;.
\eeq
We can therefore rewrite the above Lagrangian as 
\beq
{\cal L} = \int \dd^d x\, \dd^d y \Big\{ \Big[\frac{\alpha^{d/2}}{(2 \pi)^{d/2} \delta \tau} \,\ee^{-\frac{\alpha}{2} \vert \vec{z} - \vec{x} \vert^2} \Big(  \delta \tau \frac{\partial \phi}{\partial t} \Big\vert_{\vec{x}} + (\vec{z}- \vec{x}) \cdot \vec{\nabla} \phi \Big) \Big] \times \nonumber 
\eeq
\beq
\times \Big[ \frac{2 \alpha^{(d+2)/2}}{(2 \pi)^{(d+2)/2}} \,\ee^{- \frac{\alpha}{2} \vert \vec{y} - \vec{z} \vert^2}  \Big( (z^p - y^p) (\vec{z} - \vec{y}) \cdot \vec{\nabla} \phi  +  \delta \tau (z^p - y^p) \frac{\partial \phi}{\partial t} \Big\vert_y \Big)  \Big] \Big\}\,.
\eeq
Now, the terms $(\vec{x} - \vec{z}) \cdot \vec{\nabla} \phi$ and $(z^p - y^p) \partial \phi/ \partial t$ are odd with respect to $\vec{x} - \vec{z}$ and $\vec{y}- \vec{z}$ respectively, and therefore drop out of the integral. On the other hand, $(y^p - z^p)(\vec{y} - \vec{z}) \cdot \vec{\nabla} \phi$ produces terms of the form $(y^p - z^p)(y^q - z^q) \partial_q \phi$. The $p \neq q$ terms are also odd with respect to $\vec{y} - \vec{z}$ and therefore drop out as well, but the $p=q$ terms are even and, therefore, are left. These terms simplify as $(y^p - z^p)^2 \partial_p \phi$. Therefore, our new expression becomes
\beq
{\cal L} = \int \dd^d x\, \dd^d y \Big\{ \Big[\frac{ \alpha^{d/2}}{(2 \pi)^{d/2} \delta \tau} \,\ee^{-\frac{\alpha}{2} \vert \vec{z} - \vec{x} \vert^2} \times \delta \tau \frac{\partial \phi}{\partial t} \Big\vert_{\vec{x}}  \Big]  \times \Big[ \frac{2 \alpha^{(d+2)/2}}{(2 \pi)^{(d+2)/2}} \,\ee^{- \frac{\alpha}{2} \vert \vec{y} - \vec{z} \vert^2} (y^p - z^p)^2 \partial_p \phi ) \Big] \Big\},
\eeq
where it should be understood that $p \in \{1, \cdots, d \}$ is a \emph{fixed} integer, and \emph{there is no} Einstein summation convention in the above expression. By performing a simple, but rather routine, computation involving separation of variables, this expression reduces to 
\beq
{\cal L} = \frac{\partial \phi}{\partial t} \partial_p \phi \;.
\eeq
This is structurally very similar to the Lagrangian generator we started out with, which is given in Eq.\ \ref{toy1},
\beq
{\cal K} (x^{\mu}, y^{\mu}, z^{\mu}; \phi)  = [(\Gamma_{0\alpha} \phi)(x^{\mu}, z^{\mu})]  [(\Gamma_{p\alpha} \phi) (y^{\mu}, z^{\mu})]\;. \label{toy5}
\eeq
It can be checked that this similarity extends to Lagrangians with other combinations of derivatives, which makes it very easy to ``read off" the expression for ${\cal K} (x^{\mu}, y^{\mu}, z^{\mu}; \phi, A^{\mu})$ if we are given an expression for the Lagrangian density in the sought-after continuum-based theory. 

There is, however, one subtlety which leads to infinite overcountings. For example, suppose we want to ``produce" ${\cal L} = \partial_p \phi$. Naively, we can ``read off" the Lagrangian generator ${\cal K} (x^{\mu}, y^{\mu}, z^{\mu}) = (\Gamma_{p\alpha} \phi)(x^{\mu}, z^{\mu})$. This means that the integrand will be $\vec{y}$-independent. \emph{However}, we \emph{still} have to formally take an integral over $\vec{y}$. This means that we will be taking integral over a constant, thus producing infinity. This is where $I_{\alpha} (y^{\mu}, z^{\mu})$ comes along (which, by the way, is the very purpose for which it was introduced). We will incorporate the $\vec{y}$-dependence by multiplying the Lagrangian generator by $I_{\alpha} (y^{\mu}, z^{\mu})$:
\beq
{\cal L} = \partial_p f \; \Longleftrightarrow \; {\cal K} (x^{\mu}, y^{\mu}, z^{\mu}) = [(\Gamma_{p\alpha} f)(x^{\mu}, z^{\mu})][ I_{\alpha} (y^{\mu}, z^{\mu})]\;.
\eeq
It can be easily checked that the function over $y$ produced by $I_{\alpha} (k, b_k)$ will integrate to $1$ and, therefore, will not affect the result. Likewise, if we don't have any derivatives at all, we will simply put two $I_{\alpha}$-coefficients: one to ``take care" of the $\vec{x}$-dependence, and the other one to ``take care" of the $\vec{y}$-dependence. For example, the ``mass term" $m^2 \phi^2$ corresponds to a ``Lagrangian generator" according to 
\beq
{\cal L} = m^2 \phi^* \phi \; \Longleftrightarrow \; {\cal K} (x^{\mu}, y^{\mu}, z^{\mu}; \phi) = m^2  \big[\phi^* (z^{\mu}) \phi (z^{\mu})\big]\big[I_{\alpha} (x^{\mu}, z^{\mu})\big]\big[I_{\alpha} (y^{\mu}, z^{\mu})\big]\;.
\eeq 

We are finally ready to write down the Lagrangian generator for the complete Lagrangians. 
 \beq {\cal K}_{\alpha} (A^{\mu}, \phi; v^{\mu}; x, y, z)=  \nonumber \eeq
\beq = \bigg(\bigg(\frac{c_O}{c_S}-1 \bigg)  (v^{\mu} (z) v^{\rho} (z) \Gamma_{\rho; \alpha} (z, x; A^{\nu}) - v^{\nu} (z) v^{\rho} (z) \Gamma_{\rho; \alpha} (z, x, A^{\mu}) ) + \Gamma^{\mu}_{\alpha} (z, x, A^{\nu}) - \Gamma^{\nu}_{\alpha} (z,x,A^{\mu}) \bigg) \times \nonumber \eeq
\beq \times \bigg(  \bigg(\frac{c_S}{c_O} - 1 \bigg) (v_{\nu} (z) v_{\rho} (z) \Gamma_{\mu; \alpha} (z, y, A^{\rho}) - v_{\mu} (z) v_{\rho} (z) \Gamma_{\nu; \alpha} (z, y,A^{\rho})) + \Gamma_{\mu; \alpha} (z, y, A_{\nu}) - \Gamma_{\nu} (z, y, A_{\mu}) \bigg) + \nonumber \eeq
\beq + \bigg(\frac{c_O}{c_S}-1 \bigg)  v^{\mu} (z) v^{\nu} (z) \Gamma_{\nu; \alpha} (z, x; \phi^*) \Gamma_{\mu; \alpha} (z, y;  \phi) + \Gamma^{\mu}_{\alpha} (z, x; \phi^*) \Gamma_{\mu; \alpha} (z, y;  \phi) +  \label{csAlone} \eeq
\beq + e^2 I_{\alpha} (z, x) I_{\alpha} (z,y) A^{\mu} (z) A^{\nu} (z) \phi^* (z) \phi (z) v_{\mu} (z) v_{\nu} (z) \bigg(\frac{c_S}{c_O} - 1 \bigg) +  \nonumber \eeq
\beq + e^2 I_{\alpha} (z, x) I_{\alpha} (z,y) A^{\mu}(z) A_{\mu} (z)\phi^* (z)\phi (z)+ \nonumber \eeq
\beq + \ii e I_{\alpha} (z, y) A^{\mu} (z) (\phi (z) \Gamma_{\mu} (z, x, \phi^*) - \phi^* (z) \Gamma_{\mu} (z,x,\phi)) + m^2 I_{\alpha} (z, x) I_{\alpha} (z,y) \phi^2 (z) \;. \nonumber
\eeq
Here, in the expressions $\Gamma_{\mu \alpha }$, $\Gamma_{\nu \alpha }$ and $\Gamma_{\rho \alpha }$, the index $\alpha$ should \emph{not} be confused with $\mu$, $\nu$ and $\rho$. While $\mu$, $\nu$ and $\rho$ are simply Lorentzian indices, the index $\alpha$ is \emph{not}. Instead, $\alpha$ represents a ``very large" number used in $e^{-\alpha x^2/2}$. Thus, $\Gamma_{\mu \alpha } \approx \partial_{\mu}$ and $\alpha$ represents the ``degree of approximation": the larger is $\alpha$, the better is the approximation. 

\subsection*{6. Emission of signals}

In Section 2 we have defined quantum states in terms of instant communication. We have also stated, however, that we don't believe in the latter. Instead, we would like to approximate instant communication by means of superluminal signals that still move with finite, albeit very large, speed. In Sections 3-5 we have described the version of relativity we are working with that allow for such signals. Let us now proceed to describe superluminal mechanism that would produce close approximation to quantum states as described in Section 2.

We propose the following model. First, we number the LPL particles. Thus, we have LPL particles $1$ through $N$, with the $k$-th LPL particle located at $\vec{x}_k$ (which first moves according to Eq.\ (\ref{motion}) and then becomes stationary at its equilibrium position, which is assumed to have been reached by now). We attach to LPL particle number $k$ a ``clock" in the form of an oscillator $e_k (\tau)$, which evolves in time according to 
\beq
\frac{\dd^2 e_k}{\dd \tau^2} = - \omega_k^2 e_k\;, \label{Oscillation}
\eeq
where $\omega_k$ is different for each $k$ while all of them satisfy the constraint
\beq (1- \epsilon_{\omega}) \omega < \omega_k < (1+ \epsilon_{\omega}) \omega \eeq
for some common $\omega$ and "very small" constant $\epsilon_{\omega} \ll 1$; $d\tau$ is defined in terms of the metric $g_{\rm S}$ instead of $g_{\rm O}$:
\beq
\dd\tau^2 = g_{S; \mu \nu}\, \dd x^{\mu} \dd x^{\nu}\;.
\eeq
The fact that $\omega_k$ is approximately the same for each $k$ will allow the outcome to approximate Section 5 with common time slice $\delta t = 2 \pi/ \omega$, while the small difference between $\omega_k$-s will allow the order in which the signals are being emitted by different LPL particles to change over time which is necessary in order to create "randomness" that would statistically lead to the "integration" at the end. We will discuss this more in Section 8. 

Now, we also postulate the existence of a very small, but finite, constant $\epsilon_e \ll 1$. Whenever the phase of the $i$-th oscillator ``crosses" the interval $[- \epsilon_e, \epsilon_e]$ (which means that a ``clock" shows a particular time), it sends a pulse with ``superluminal" (but finite) speed $c_{\rm S} \gg c_{\rm O}$. In light of the fact that $\epsilon_e$ is small, that pulse has a very short duration. During the passage of that pulse, other LPL particles perform some simple "steps" in "calculation" that happen to be functions of the parameters of the particle that emitted the pulse; after several "rounds" of different particles emitting pulses and different "basic steps" being performed, we would obtain the desired values of $\psi_k$ in the emergent limit. 

Now, in order for the pulses to accomplish the desired effect, they have to carry the information about the internal parameters of the LPL particle that sent these pulses. Thus, we formally introduce ``image fields" $I_q (x)$, $\vec{I_x} (x)$, $I_A^{\mu} ( (x)$, $I_{\phi} (x)$ and $I_{\psi} (x)$ whose values approximate $q_i$, $\vec{x}-\vec{x}_i$, $A_i^{\mu}$, $\phi_i$ and $\psi_i$ respectively. In covariant notation, this can be formulated as 
\beq
(I_q, I_x^{\mu}, I_A^{\mu}, I_{\phi}, I_{\psi})  (x^{\mu}) \approx 
\left\{
	\begin{array}{ll}
		(q_j, x^{\mu}_{\perp V}-x^{\mu}_{i \perp V}, A_j^{\mu}, \phi_j, \psi_j)  & {\rm if\ dot}\ j\ {\rm emitted\ the\ signal\ reaching}\ \vec{x}\ {\rm at\ time}\ t\\
		(0, 0, 0, 0, 0) & {\rm otherwise}.
	\end{array}
\right. \label{QXpsi2}
\eeq
where
\beq
b^{\mu}_{\parallel a} = \frac{a^{\mu} a^{\nu}b_{\nu}}{a^{\rho}a_{\rho}} \; , \; b^{\mu}_{\perp a} = b^{\mu} - \frac{a^{\mu} a^{\nu}b_{\nu}}{a^{\rho}a_{\rho}}\;.
\eeq
Thus, at the equilibrium situation, 
\beq V^{\mu} \approx \delta^{\mu}_0 \Rightarrow x^{\mu}_{\perp V} = \vec{x} \eeq
In light of the ``classical" nature of the desired theory, Eq.\ (\ref{QXpsi2}) cannot be simply postulated. Instead, we want to come up with a set of differential equations that produces the latter. In fact, we would like the definitions of $I_q$, $I_x^{\mu}$, $I_A^{\mu}$, $I_{\phi}$ and $I_{\psi}$ per the sought-after differential equations to be exact, while Eq.\ (\ref{QXpsi2}) will be an emergent approximation. As we mentioned previously, the ``source" of a signal is an LPL particle whose oscillator crosses the region $[- \epsilon_e, \epsilon_e]$. Thus, the source term needs to include a ``conditional" function. For this purpose, we will define the ``truth value" of a statement as follows:
\beq
T({\rm true}) = 1 \; ; \quad T({\rm false}) = 0\;.
\eeq
In order to accommodate a reader who ``for philosophical reasons" wants everything to be differentiable whenever possible, we will introduce a differentiable approximation to the above given ``truth value":
\beq
T_{\epsilon_T} (x \in [a,b]) = \Big( \frac{1}{2} + \frac{2}{\pi} \tan^{-1} \frac{x-a}{\epsilon_T} \Big) \Big( \frac{1}{2} + \frac{2}{\pi} \tan^{-1} \frac{b-x}{\epsilon_T} \Big)\,. \label{T1}
\eeq
The wave operator for the wave propagating at the speed $c_{\rm S}$ is $g_{\rm S}^{\mu\nu} \,\partial_{\mu} \partial_{\nu}$. The signals propagating under this operator would decrease their strength over distance, contrary to Eq.\ \ref{QXpsi2}. Therefore, we will introduce two sets of fields: $\mu$-s and $I$-s. The $\mu$-fields will attenuate with the distance, while $I$-fields will not. The analysis of $\mu$-fields will make it possible to deduce the distance to the source (based on ``local" information alone) and, therefore, ``manufacture" non-attenuating expressions, which will be identified with $I$-s.

Let us now write it more explicitly. The $\mu$-fields will behave according to the wave equation: 
\beq
g_{\rm S}^{\alpha \beta} \partial_{\alpha} \partial_{\beta} (\mu_q, \mu_{\psi}, \mu_{\cal L}, \mu_S, \mu_{\phi}, \mu_A^{\mu})   - \epsilon_d g_{\rm S}^{\alpha \beta} V_{\alpha} \partial_{\beta} (\mu_q, \mu_{\psi}, \mu_{\cal L}, \mu_S, \mu_{\phi}, \mu_A^{\mu})+ m_{\mu}^2 (\mu_q, \mu_{\psi}, \mu_{\cal L}, \mu_S, \mu_{\phi}, \mu_A^{\mu})= \nonumber \eeq
\beq = \sum_{j=1}^N \Big( (q_j, \psi_j, {\cal L}_j,  S_j, \phi_j, A_j^{\mu}) \int_{\gamma_j}\dd\tau\, \delta^4 (x - \gamma_j (\tau))\, T_{\epsilon_T} \Big(\frac{1}{e_j} \frac{\dd e_j}{\dd\tau} \in [- \epsilon_e, \epsilon_e] \Big) \Big)  \label{Messengers1}\eeq
where the purpose of the $T$s is to make sure that the signals are emitted in pulses, and the duration of the pulses is limited by the time period where the ``conditions" of the $T$s are satisfied.  As before, the purpose of the $\epsilon_d$-term is to provide a very small friction coefficient that would be unnoticeable within the same lattice but prevent communication across different lattices. And, again as was mentioned earlier, the $V_{\alpha}$ in the $\epsilon_d$-term is basically a $c_{\rm S}$-covariant replacement of $\delta_{\alpha}^0$ which is guaranteed to be timelike and positive due to the initial conditions (\ref{Initial}).

If we notice that the signal has a finite duration and we assume that it has ``started" some time ago and will continue on for some more time, we can use the spherical symmetry in $\mathbb{R}^3$ to say that $\mu_q$ is inversely proportional to the projected distance on a spacelike hypersurface, $\vert x_{\perp V} - \gamma_{\perp V} (\tau) \vert$, to an approximation of the order of a function of  the ``friction term" $\epsilon_d$. Therefore, it is easy to check that the desired conditions for $I_x^{\alpha}$, $I_{\psi}$, $I_q$, $I_{\cal L}$, $I_S$, $I_{\phi}$ and $I_A^{\mu}$  are approximately met during the majority of the time of a pulse if we define these according to
\beq
I_x^{\alpha} = \frac{\ee^{- \kappa/ \mu_q} \mu_q g_{\rm S}^{\alpha \beta} \partial_{\perp v; \beta} \mu_q}{g_{\rm S}^{\gamma \delta} \partial_{\perp v; \gamma} \mu_q \partial_{\perp v; \delta} \mu_q }  \eeq

\beq \quad (I_{\psi}, I_q, I_{\cal L}, I_S, I_{\phi}, I_A^{\mu})  = \frac{\ee^{- \kappa/ \mu_q} \mu_q (\mu_{\psi}, \mu_q, \mu_{\cal L}, \mu_S, \mu_{\phi}, \mu_A^{\mu})}{\sqrt{g_{\rm S}^{\alpha \beta} \partial_{\alpha} \mu_q \partial_{\beta} \mu_q}}  \;; \quad 
\nonumber \eeq
Here, $\kappa$ is very small; thus, throughout the duration of the signal, $e^{- \kappa/ \mu_q} \approx 1$ which is what we need to assume in order for the first line on the right-hand side of Eq.\ (\ref{QXpsi2}) to be satisfied. On the other hand, whenever $\mu_q$ is very small, we get $e^{- \kappa/ \mu_q} \approx 0$. In fact, $e^{- \kappa/ \mu_q}$ goes to zero much faster than anything else that might go to infinity, which is why we obtain $I_x^{\alpha} \approx I_{\psi} \approx I_q \approx 0$, which means that the second line of Eq.\ (\ref{QXpsi2}) is satisfied as well.

\subsection*{7. Generic processing of signals}

So far we have described a mechanism through which the information about the LPL particle \emph{emitting} a signal is available throughout space. We would now like this information to be processed by other LPL particles. Thus, once the duration of the short pulse (which was propagating through the continuum) is over, the information that the pulse has carried has been "stored" \emph{inside LPL particles}. It is important to note, however, that any given LPL particle has very limited "memory". After all, part of the reason we want to get rid of configuration space on the first place is that it has too many coordinates which makes it unphysical. Thus, the "physical" quantities (such as LPL particles) should have only few parameters. After all, each of such parameters can be viewed as a "discretization" of "classical fields" (a subset of these fields will correspond to known fields from conventional physics; others will need to be added in order to "aid" the "memory mechanism" of LPL particles; the latter fields would not correspond to anything conventional and can not be detected in the lab, but they are still viewed as "additional \emph{physical} fields" nonetheless). Even though there will be more "fields" than the ones that have counterparts in conventional physics, we would still like their number to be small in order for the "principle of it" to look "classical". Since the "memory" of LPL particles is encoded by these parameters, the memory of LPL particle should be limited. This means that LPL particle can not possibly "remember" \emph{all} of the other LPL particles that have emitted a signal. Instead, it "remembers" the last three. 

If the last three particles that emitted the signal were $i$, $j$ and $k$, then particle number $l$ "remembers" the information about $(\psi_i, q_i, {\cal L}_i, S_i, \phi_i, A_i^{\mu})$, $(\psi_j, q_j, {\cal L}_j, S_j, \phi_j, A_j^{\mu})$ and $(\psi_k, q_k, {\cal L}_k, S_k, \phi_k, A_k^{\mu})$ in a form of $(\psi_{l1}, q_{l1}, {\cal L}_{l1}, S_{l1}, \phi_{l1}, A_{l1}^{\mu})$, $(\psi_{l2}, q_{l2}, {\cal L}_{l2}, S_{l2}, \phi_{l2}, A_{l2}^{\mu})$ and $(\psi_{l3}, q_{l3}, {\cal L}_{l3}, S_{l3}, \phi_{l3}, A_{l3}^{\mu})$, respectively (which are not to be confused with $(\psi_l, q_l, {\cal L}_l, S_l, \phi_l, A_l^{\mu})$). It then uses the above information to perform a sequence of steps to modify ${\cal L}_l$, $S_l$ and $\psi_l$. Needless to say, since it only "knows" the information about three other LPL particles this can not possibly be sufficient to reproduce "integrals" over "all" LPL particles. \emph{Nevertheless}, as the new signals get emitted, the LPL particle $l$ keeps "forgetting" some of the particles and instead "remembers" others. Thus, it is conceivable that statistically we would, indeed, expect the "integrals" to emerge. Our goal is to design the steps in such a way that it happens. 

Now, as will be seen below, during the time the LPL particle receives the signal, it performs more than just one step; this is possible due to the fact that the duration of a signal is still finite, despite the fact that it is very small. Thus, different steps can still be separated in time. This raises the question: how can it "know" how many steps it has already completed and which step it is "in the process of" performing, without violating locality in time? We propose to divide signal into $n$ different "parts". The signal (emitted by the outside particle) will carry extra parameter, $I_P (\vec{x},t)$ ("P" stands for "part") and during "part P" we have $I_P (\vec{x},t) \approx P$. Thus, the LPL particle $l$ doesn't remember what it did before. It simply reads off the value $I_P (x_l^{\mu})$ and performs an "infinitesimal part" of the "step number P" where $ I_P (x_k^{\mu}) \approx P \in \mathbb{N}$. Thus, we would like $I_P (x^{\mu})$ to look like a continuous approximation of step function in $x^0$. As usual, we accomplish it by first introducing $\mu_P$; and we make its source resemble step function we would like to see: 
\beq g_S^{\alpha \beta} \partial_{\alpha} \partial_{\beta} \mu_P - \epsilon_d g_S^{\alpha \beta} V_{\alpha} \partial_{\beta} \mu_P = \sum_{j=1}^n j \; T_{\epsilon_T} \Bigg(\frac{1}{e_j} \frac{de_j}{d \tau} \in \Bigg[- \epsilon_e + \frac{2 \epsilon_e (j-1)}{n},  \frac{2 \epsilon_e j}{n} \Bigg] \Bigg) \eeq
Then we use our usual prescription to "convert" $\mu_P$ into $I_P$, 
\beq \quad I_P  = \frac{\ee^{- \kappa/ \mu_q} \mu_q \mu_P}{\sqrt{g_{\rm S}^{\alpha \beta} \partial_{\alpha} \mu_q \partial_{\beta} \mu_q}}  \;; \quad 
\nonumber \eeq
Now, as a toy model, suppose we were to have three parameters attached to the LPL particle $l$, namely, $a_l$, $b_l$, and $C_l^{\mu}$. And suppose we wanted to reproduce the following algorithm,

a) $a_l \longleftarrow f (a_l, b_l, C_l^{\mu})$

b) $b_l \longleftarrow g (a_l, b_l, C_l^{\mu})$

c) $C_l^{\mu} \longleftarrow w_3^{\mu} (a_l, b_l, C_l^{\mu})$

d) $a_l \longleftarrow h (a_l, b_l, C_l^{\mu}) $

e) $b_l \longleftarrow i (a_l, b_l, C_l^{\mu})$

Now, in light of the fact that each of these steps has continuous dynamics, it has to take finite amount of time. So when we are "in the middle" of step a, we no longer have "old" value of $a_l$ nor do we have "new" one yet. Rather, it has some "intermediate" value. However, we would like to use $f$ evaluated for the "old" value of $a_l$ in order to determine what "new" value to "aim for". This means that we would like to "store" the "old" value of $f$ inside some parameter $f_l$ so that even when "old data" is no longer available we still "remember" the value of $f$ when it was. Then we use that stored value of $f$ as an aiming point, \emph{instead of} computing its running value throughout the process. In other words, part "a" will be replaced with 

a') $f_l \longleftarrow f (a_l, b_l, C_l^{\mu})$

a'') $a_l \longleftarrow f_l$

By making similar changes in the rest of steps, we have

b') $f_l \longleftarrow g (a_l, b_l, C_l^{\mu})$

b'') $b_l \longleftarrow f_l$

c') $w_l^{\mu} \longleftarrow w_3^{\mu} (a_l, b_l, C_l^{\mu})$

c'') $C_l^{\mu} \longleftarrow w_l^{\mu}$

d') $f_l \longleftarrow h (a_l, b_l, C_l^{\mu})$

d'') $a_l \longleftarrow f_l$

e') $f_l \longleftarrow i (a_l, b_l, C_l^{\mu})$

e'') $b_l \longleftarrow f_l$

Thus, we have 10 steps, and the time allocated to complete each of them is $2 \epsilon_e /n$. Now, a generic $\alpha \rightarrow \beta$ can be generated through
\beq [\alpha \rightarrow \beta] \Leftrightarrow \bigg[ \epsilon_{\rightarrow} \frac{d \alpha}{d \tau} = \beta - \alpha \bigg] \; , \; \epsilon_{\rightarrow} \ll \epsilon_e \eeq
Furthermore, part number $P$ in the algorithm is indicated by $I_P(x_k^{\mu}) \approx P$, where approximation sign is due to the fact that $P$ is an integer while $I_P (x_k^{\mu})$ is not (thanks to the "continuum" dynamics of the latter). If the criteria of approximation is too strict, we run the risk that "not enough time" is spent where it is satisfied and the desired process wont be completted. On the other hand, if it is too lose, then "too much time" wont do any harm since, once the "aimed-for" value is "almost reached" the subsequent dymanics wont cause significant changes. Because of this, we will make the losest possible criteria for approximation; namely, 
\beq {\rm Part \; P } \Longleftrightarrow I_P (x_l^{\mu}) \approx P \Longleftrightarrow I_P (x_l^{\mu}) \in [P-1/2, P+1/2] \eeq
Therefore,
\beq [{\rm Part \; P: \alpha_l \rightarrow \beta_l}] \Longrightarrow \bigg[ \epsilon_{\rightarrow} \frac{d \alpha}{d \tau} = (- \alpha + \beta) T_{\epsilon_T} \bigg(I_P (x_l^{\mu}) \in \bigg[P-\frac{1}{2}; P+ \frac{1}{2} \bigg] \bigg) \bigg] + \cdots \eeq
This means that the above a'-e'' translate to
\beq \epsilon_{\rightarrow} \frac{d f_l}{d \tau} = (- f_l + f (a_l, b_l, C_l^{\mu})) T_{\epsilon_T} (I_P (x_l^{\mu}) \in [1/2, 3/2]) + \nonumber \eeq
\beq + (-f_l +g (a_l, b_l, C_l^{\mu})) T_{\epsilon_T} (I_P (x_l^{\mu}) \in [5/2,7/2]) +  \eeq
\beq +  (-f_l+h (a_l, b_l, C_l^{\mu})) T_{\epsilon_T} (I_p (x_l^{\mu})\in [13/2, 15/2]) +  \nonumber \eeq
\beq + (-f_l+ i (a_l, b_l, C_l^{\mu})) T_{\epsilon_T} (I_p (x_l^{\mu}) \in [17/2, 19/2]) \nonumber  \eeq
\beq \epsilon_{\rightarrow} \frac{d a_l}{d \tau} = (-a_l + f_l) (T_{\epsilon_T} (I_p (x_l^{\mu})  \in [3/2,5/2]) + T_{\epsilon_T} (I_p (x_l^{\mu}) \in [15/2,17/2])) \eeq
\beq \epsilon_{\rightarrow} \frac{d b_k}{d \tau} = (-b_l + f_l) (T_{\epsilon_T} (I_P(x_l^{\mu}) \in [7/2,9/2]) + T_{\epsilon_T} (I_P(x_l^{\mu}) \in [19/2, 21/2])) \eeq
\beq \epsilon_{\rightarrow}\frac{d C_l^{\mu}}{d \tau} = (-C_l^{\mu} + w_l^{\mu}) T_{\epsilon_T} (I_P(x_l^{\mu}) \in [11/2,13/2]) \eeq
The above can be generalised to the process of any other number of steps. In other words we have established a one-to-one correspondence between the desired step-by-step process and the differential equations that would lead to the emergence of said process.

\subsection*{8. Specific algorithm}

Let us now come up with specific algorithm that produces discretization of QFT described in Section 5. It is easy to see that the path integral can be computed "slice by slice" according to 
\beq \psi (\phi_{(q)}, A^{\mu}_{(q)}, a \delta t) = \sum_{q'} \psi (\phi_{(q')}, A^{\mu}_{(q')}, (a-1) \delta t) \exp \Bigg(i \delta t \;  \int {\cal K} d^3 x d^3 y d^3 z (x, y, z, \phi_{(q')}, A^{\mu}_{(q')}) \Bigg) \label{NonlocalEvolution} \eeq
Thus, the "integral" in the exponent, 
\beq {\cal L} (\vec{z}, a \delta t; \phi, A^{\mu}) = \int d^3 x d^3 y {\cal K}  (x, y, z, \phi_{(q')}, A^{\mu}_{(q')}) \label{DesiredL}\eeq
is, per inspection of right hand side, a function of $(q, q'; \vec{z}, t)$, as compared to $\psi$ which is a function of $(q, t)$. Now, in order for "encoding" of $\psi$ to be consistent we had to assume
\beq q_i = q_j \Longrightarrow \psi_i (t) \approx \psi_j (t) \label{psiConsistent} \eeq
For $\cal L$ we have to modify this constraint as 
\beq [(q_i, q'_i) = (q_j, q'_j) \wedge \vec{x}_i \approx \vec{x}_j] \Longrightarrow {\cal L}_i (t) \approx {\cal L}_j (t) \eeq
Finally,  the action
\beq S = \int d^3 z {\cal L} (\vec{z}, t) \eeq
is a function of $(q,q')$; thus the consistency requires that we have a condition 
\beq (q_i, q'_i)=(q_j,q'_j) \Longrightarrow S_i (t) = S_j (t) \eeq
While the same-time correlations are different for $\psi$, $\cal L$ and $S$, in all three cases the values are pointwise recorded in a form of ${\cal L}_i$, $S_i$, and $\psi_i$. Now, $\cal L$ is an integral of $\cal K$ and $S$ is an integral of $\cal L$. Thus, we need an algorithm to evaluate sum. As a "toy model" consider the following sequence, 
\beq A_{k+1} = (1 - \epsilon_{A1}) A_k + \epsilon_{A2} a_k \; , \; a_k \in \{b_1, \cdots, b_p \} \label{SequenceExample} \eeq
for some aforegiven $B= \{b_1, \cdots, b_p \}$. It can be shown by induction that 
\beq A_n = (1- \epsilon_{A1})^{n-m} A_m + \epsilon_{A2} \sum_{k=m}^{n-1} (1- \epsilon_{A1})^{n-1-k} b_k \label{InductionSum1} \eeq
If we assume $n-m \gg p$, the first term on right hand side will become negligible; furthermore, the term containing $b_l$ would appear approximately $(n-m)/p$ times in the sum. Thus, we obtain
\beq A_n \approx \epsilon_{A2} \sum_{l=1}^p \bigg(b_l \sum_{k=1}^{[ (n-m)/p]} (1- \epsilon_{A1})^{pk}  \bigg) \approx \frac{\epsilon_{A2}}{\epsilon_{A1}} (1- (1- \epsilon_{A1})^{(n-m)/p}) \sum_{l=1}^p b_l  \approx \frac{\epsilon_{A2}}{\epsilon_{A1}}  \sum_{l=1}^p b_l \eeq
which produces desired result if we set $\epsilon_{A1} = \epsilon_{A2}$. 

Let us now return to ${\cal L}_p$. Our first goal is to "generate" Eq \ref{DesiredL}. Thus, $\{b_1, \cdots, b_p \}$ is replaced with $\{ {\cal K} (x_i, x_j, x_k, \phi_i, A^{\mu}_i, \phi_j, A^{\mu}_j, \phi_k, A^{\mu}_k) \vert \vert x_k - x_l \vert < \epsilon_x \}$. Thus, at any given time, point $l$ adds \emph{one} of these $\cal K$-s per prescription in Eq \ref{SequenceExample}. Now, since we would like our theory to be "local" we are "not allowed" to "directly" use the information pertaining to other particles. Instead, we use the "record" of other particles "within" particle $l$. As was stated in the beginning of Section 7, particle $l$ only has enough degrees of freedom to record information about three other particles at any given time. Since, over time, LPL particle $l$ changes around the other three particles it had stored, we would statistically expect different $\cal K$-s to be added at random, just like Eq \ref{SequenceExample} prescribes. This is true due to the fact that $\omega_k$ different for each $k$ which allows the sequence of emission of signals to change over time and, as a result, all possible "triplets" will appear at some time or the other.  Now, the sum of \ref{SequenceExample} is produced per the following algorithm:

{\bf Rule 1:} If $x^{\mu}_{3p} < \Delta x$ then

$ {\cal L}_l \longleftarrow (1- \epsilon_{\cal K}) {\cal L}_l + \epsilon_{\cal K} (\delta v) {\cal K} (\phi_{1l}, A^{\mu}_{1l}, x^{\mu}_{1l}, \phi_{2l}, A^{\mu}_{2l}, x^{\mu}_{2l}, \phi_{3l}, A^{\mu}_{3l}, x^{\mu}_{3l})$. 

In order for the repetitions of the "Rule 1" to cover a "random sample" of $\cal K$-s, we need to alternate the values of the variables inside of $\cal K$ (equivalently, we are alternating the choice of triple of "other" particles that particle $l$ "knows about"). Whenever new LPL particle emits a signal, the information contains in that signal replaces the previous values of some of the internal parameters of LPL particle $l$. Now, we need to consider two different cases, $q=q'_l$ (extra indexes $1$ and $2$) and $q=q_l$ (extra index $3$). The former replacement will be done per Rule 2 and the latter per Rule 3:  

{\bf Rule 2:} If $I_q (x^{\mu}_l) = q'_l$ then 

a) $ (q_{1p}, q'_{1l}, {\cal L}_{1l}, S_{1l}, \psi_{1l}, \phi_{1l}, A^{\mu}_{1l}, x^{\mu}_{1l}) \longleftarrow (q_{2l}, q'_{2l}, {\cal L}_{2l}, S_{2l}, \psi_{2l}, \phi_{2l}, A^{\mu}_{2l}, x^{\mu}_{2l}) $

b) $(q_{2l}, q'_{2l}, {\cal L}_{2l}, S_{2l}, \psi_{2l}, \phi_{2l}, A^{\mu}_{2l}, x^{\mu}_{2l}) \longleftarrow$

$\longleftarrow (I_q (x^{\mu}_l), I_q' (x^{\mu}_l), I_{\cal L} (x^{\mu}_l), I_S (x^{\mu}_l), I_{\psi} (x^{\mu}_l), I_{\phi} (x^{\mu}_l), I_A^{\mu} (x^{\mu}_l), I_x^{\mu} (x^{\mu}_l)) $ 

{\bf Rule 3:} If $I_q (\vec{x}_l,t) \approx q_l$ then 

$ (q_{3l}, q'_{3l}, {\cal L}_{3l}, S_{3l}, \psi_{3l}, \phi_{3l}, A^{\mu}_{3l}, x^{\mu}_{3l})  \longleftarrow$

$\longleftarrow  (I_q (x^{\mu}_l), I_q' (x^{\mu}_l), I_{\cal L} (x^{\mu}_l), I_S (x^{\mu}_l), I_{\psi} (x^{\mu}_l), I_{\phi} (x^{\mu}_l), I_A^{\mu} (x^{\mu}), I_x^{\mu} (x^{\mu}_l))$

We then "integrate" $\cal L$ to obtain $S$ per similar strategy to the one used in "rule 1":  

{\bf Rule 4:} If $I_q (x_l^{\mu}) \approx q_l$ and $I_q' (x_l^{\mu}) = q'_l$ then $S_l \longleftarrow (1 - \epsilon_{\cal L}) S_l + \epsilon_{\cal L} (\Delta v) I_{\cal L} (x^{\mu}_l)$.

Now that we have "encoded" the action, it is time go go back to the evolution of $\psi$. First of all, in order for $\psi$ to be consistent, it has to satisfy Eq \ref{psiConsistent}. The easiest way to do it is to make sure that whenever a particle $k$ emits a signal, the particle $l$ "copies" the value of $\psi$ from the particle $k$ as long as $q_k=q_l$. The value of $\psi_k$ is not "locally" available at the location of particle $l$; instead, particle $l$ "knows" about it by measuring $I_{\psi} (x^{\mu}_l)$ which, at the moment, approximates $\psi_k$. Thus, we postulate 

{\bf Rule 5:} If $I_q (x^{\mu}_l) \approx q_l$ then $\psi_l \longleftarrow I_{\psi} (x^{\mu}_l)$

Now, we would like the evolution of $\psi$ to reproduce path integral. If we go back to the language of functionals, it is easy to see, by induction, that path integral is reproduced if
\beq \psi (\phi_{(q)}) = \sum_{q'} \psi (\phi_{(q')}) S (q',q, \cdots) \label{Induction} \eeq
Now, in order to perform "one step" of the above sum, we can instruct a given point $l$ to "add" one single term of the above sum whenever it receives a signal from any particle whose value of $q'$ matches "un-primed" $q_l$ (the reason it has to be done during the passage of a signal is that $\psi_l$ on the right hand side of Eq \ref{Induction} varies, and particle $p$ can only "know" its current value by reading off the value of $I_{\psi}$ from the signal):

{\bf Rule 6:} If $I_q (x^{\mu}_l) \approx q'_l$ then $\psi_l \longleftarrow I_{\psi} (x^{\mu}_l) e^{i \delta t I_S}$

However, in order to actually arrive at the sum, Rule 6 needs to be performed several times. But, due to the fact that Rule 5 is repeatedly performed as well, most of the performances of Rule 6 will get "erased" through 5 before they have a chance to be "added" to newer performances. In fact, the effect of Rule 6 can "survive" only when LPL particle $l$ emits a signal \emph{before} any other LPL particle does with the same $q$ (although it is fine if a point with $q$ other than $q_l$ emits a signal before $l$ since in this case its signal is ignored per rule 15). Now there are approximately $N/M$ particles with $q=q_l$ is $N/M$ which makes the probability of the above extremely small. But on the scales
\beq \Delta t \gg \frac{N \delta t}{M} \eeq
the "unlikely" events get a chance to accumulate which results in the emergence of expected time evolution, with rescaled time. 

Now, we recall that in the previous section we were splitting each rule (such as a) into two rules (such as a' and a''). We will do the same thing here. Additionally, we will replace $r \approx n$ with $n-1/2<r<n+1/2$ (for the same reasons as given in previous section) and also all the "if" statements with corresponding $T_{\epsilon_T}$-s. This leads to the following set of rules:

1') $f_l \longleftarrow {\cal L}_l - \epsilon_{\cal K} [ {\cal L}_l - (\delta v) {\cal K} (\phi_{1l}, A^{\mu}_{1l}, x^{\mu}_{1l}, \phi_{2l}, A^{\mu}_{2l}, x^{\mu}_{2l}, \phi_{3l}, A^{\mu}_{3l}, x^{\mu}_{3l})] \times$

$\times T_{\epsilon_T} (I_x^{\mu} (x^{\mu}_{3l}) I_{x \mu} (x^{\mu}_{3l}) \in (- (\Delta x)^2, 0))$

1'') ${\cal L}_l \longleftarrow f_l$

2a') $(q_{4l}, q'_{4l}, {\cal L}_{4l}, S_{4l}, \psi_{4l}, \phi_{4l}, A^{\mu}_{4l}, x^{\mu}_{4l}) \longleftarrow $

$\longleftarrow (q_{2l}, q'_{2l}, {\cal L}_{2l}, S_{2l}, \psi_{2l}, \phi_{2l}, A^{\mu}_{2l}, x^{\mu}_{2l}) T(I_q (x^{\mu}_l) \in (q'_l-1/2, q'_l+1/2)) + $

$+ (q_{1l}, q'_{1l}, {\cal L}_{1l}, \psi_{1l}, \phi_{1l}, A^{\mu}_{1l}, x^{\mu}_{1l}) (1-  T(I_q (x^{\mu}_l) \in (q'_l-1/2, q'_l+1/2)))$

2a'') $(q_{1l}, q'_{1l}, {\cal L}_{1l}, \psi_{1l}, \phi_{1l}, A^{\mu}_{1l}, x^{\mu}_{1l}) \longleftarrow (q_{4l}, q'_{4l}, {\cal L}_{4l}, S_{4l}, \psi_{4l}, \phi_{4l}, A^{\mu}_{4l}, x^{\mu}_{4l})$

2b') $(q_{4l}, q'_{4l}, {\cal L}_{4l}, S_{4l}, \psi_{4l}, \phi_{4l}, A^{\mu}_{4l}, x^{\mu}_{4p}) \longleftarrow$

$\longleftarrow (I_q (x^{\mu}_l), I_q' (x^{\mu}_l), I_{\cal L} (x^{\mu}_l), I_S (x^{\mu}_l), I_{\psi} (x^{\mu}_l), I_{\phi} (x^{\mu}_l), I_A^{\mu} (x^{\mu}_l), I_x^{\mu} (x^{\mu}_l)) \times$ 

$\times T(I_q (x^{\mu}_l, t) \in (q'_l-1/2, q'_l+1/2)) + $

$+  (q_{2l}, q'_{2l}, {\cal L}_{2l}, S_{2l}, \psi_{2l}, \phi_{2l}, A^{\mu}_{2l}, x^{\mu}_{2l}) (1- T(I_q (x^{\mu}_l, t) \in (q'_l-1/2, q'_l+1/2)) )$ 

2b'') $(q_{2l}, q'_{2l}, {\cal L}_{2l}, S_{2l}, \psi_{2l}, \phi_{2l}, A^{\mu}_{2l}, x^{\mu}_{2l}) \longleftarrow (q_{4l}, q'_{4l}, {\cal L}_{4l}, S_{4l}, \psi_{4l}, \phi_{4l}, A^{\mu}_{4l}, x^{\mu}_{4l})$ 

3')  $ (q_{4l}, q'_{4l}, {\cal L}_{4l}, S_{4l}, \psi_{4l}, \phi_{4l}, A^{\mu}_{4l}, x^{\mu}_{4l}) \longleftarrow$

$\longleftarrow  (I_q (\vec{x}_l), I_q' (x^{\mu}_l), I_{\cal L} (x^{\mu}_l), I_S (x^{\mu}_l), I_{\psi} (x^{\mu}_l), I_{\phi} (x^{\mu}_l), I_A^{\mu} (x^{\mu}_l),  I_x^{\mu} (x^{\mu}_l)) \times$

$\times T_{\epsilon_T} (I_q (x^{\mu}_l, t) \in (q_l-1/2, q_l+1/2)) + $

$+  (q_{3l}, q'_{3l}, {\cal L}_{3l}, S_{3l}, \psi_{3l}, \phi_{3l}, A^{\mu}_{3l}, x^{\mu}_{3l}) (1-T_{\epsilon_T} (I_q (x^{\mu}_l, t) \in (q_l-1/2, q_l+1/2))) $

3'')  $ (q_{3l}, q'_{3l}, {\cal L}_{3l}, S_{3l}, \psi_{3l}, \phi_{3l}, A^{\mu}_{3l}, x^{\mu}_{3l}) \longleftarrow  (q_{4l}, q'_{4l}, {\cal L}_{4l}, S_{4l}, \psi_{4l}, \phi_{4l}, A^{\mu}_{4l}, x^{\mu}_{4l}) $

4') $f_l \longleftarrow S_l - \epsilon_{\cal L} T(I_q (x^{\mu}_l) \in (q_l-1/2,q_l+1/2)) \times$

$\times  T_{\epsilon_T} (I_q' (x^{\mu}_l) \in (q'_l-1/2,q'_l+1/2)) ( S_l - (\Delta v) I_{\cal L} (x^{\mu}_l))$.

4'') $S_l \longleftarrow f_l$

5', 6') $f_l \longleftarrow I_{\psi} (\vec{x}_l) (T(I_q (\vec{x}_l) \in (q_l-1/2, q_l+1/2)) + e^{i \delta t I_S (x^{\mu}_l)} T_{\epsilon_T} (I_q (x^{\mu}_l) \in (q'_l - 1/2, q'_l+1/2))$

5'', 6'') $\psi_l \longleftarrow f_l$

Finally we mimic the strategy of Section 7 to convert the above algorithm into differential equations:

\beq \epsilon_{\rightarrow} \frac{d f_l}{d \tau} = \{- f_l +  {\cal L}_l - \epsilon_{\cal K}  T_{\epsilon_T} [I_x^{\mu} (x^{\mu}_{3l}) I_{x \mu} (x^{\mu}_{3l}) \in (- (\Delta x)^2, 0)] \times \nonumber \eeq
\beq \times  [ {\cal L}_l - (\delta v) {\cal K} (\phi_{1l}, A^{\mu}_{1l}, x^{\mu}_{1l}, \phi_{2l}, A^{\mu}_{2l}, x^{\mu}_{2l}, \phi_{3l}, A^{\mu}_{3l}, x^{\mu}_{3l})] \} T_{\epsilon_T} [I_P (x_l^{\mu}) \in (1/2, 3/2) ] + \nonumber \eeq
\beq + \{-f_l + S_l - \epsilon_{\cal L} T_{\epsilon_T}[I_q (x^{\mu}_l) \in (q_l-1/2,q_l+1/2)]  T_{\epsilon_T} [I_q' (x^{\mu}_l) \in (q'_l-1/2,q'_l+1/2)] \times \nonumber \eeq
\beq \times [ S_l - (\Delta v) I_{\cal L} (x^{\mu}_l)]\} T_{\epsilon_T} [I_P (x_l^{\mu}) \in (17/2,19/2)] +  \eeq
\beq +  \big\{-f_l+ I_{\psi} (x^{\mu}_l) \{T_{\epsilon_T}[I_q (x^{\mu}_l) \in (q_l-1/2, q_l+1/2)] + \nonumber \eeq
\beq + e^{i \delta t I_S (x^{\mu}_l)} T_{\epsilon_T} [I_q (x^{\mu}_l) \in (q'_l - 1/2, q'_l+1/2)]\} \big\} T_{\epsilon_T} [I_P (x_l^{\mu})\in (21/2, 23/2)]  \nonumber \eeq
\beq \epsilon_{\rightarrow} \frac{d {\cal L}_l}{d \tau} = (-{\cal L}_l + f_l) T_{\epsilon_T} [I_P (x_l^{\mu})  \in (3/2,5/2)]  \eeq
\beq \epsilon_{\rightarrow} \frac{d S_l}{d \tau} = (-S_l + f_l) T_{\epsilon_T} [I_P(x_l^{\mu}) \in (15/2,17/2)]  \eeq 

\beq \epsilon_{\rightarrow}\frac{d (q_{4l}, q'_{4l}, {\cal L}_{4l}, S_{4l}, \psi_{4l}, \phi_{4l}, A^{\mu}_{4l}, x^{\mu}_{4l}) }{d \tau} = \big\{ - (q_{4l}, q'_{4l}, {\cal L}_{4l}, S_{4l}, \psi_{4l}, \phi_{4l}, A^{\mu}_{4l}, x^{\mu}_{4l})  + \nonumber \eeq
\beq + T_{\epsilon_T}[I_q (x^{\mu}_l, t) \in (q'_l-1/2, q'_l+1/2)]  (q_{2l}, q'_{2l}, {\cal L}_{2l}, S_{2l}, \psi_{2l}, \phi_{2l}, A^{\mu}_{2l}, x^{\mu}_{2l})+ \nonumber \eeq
\beq + \{1-  T_{\epsilon_T}[I_q (x^{\mu}_l, t) \in (q'_l-1/2, q'_l+1/2)]\} \times \nonumber \eeq
\beq \times  (q_{1l}, q'_{1l}, {\cal L}_{1l}, \psi_{1l}, \phi_{1l}, A^{\mu}_{1l}, x^{\mu}_{1l}) \big\}  T_{\epsilon_T} [I_P (x_l^{\mu}) \in (5/2, 7/2) ]+ \nonumber \eeq
\beq + \big\{-f_l +  T_{\epsilon_T}[I_q (x^{\mu}_l, t) \in (q'_l-1/2, q'_l+1/2)] \times \nonumber \eeq
\beq \times (I_q (x^{\mu}_l), I_q' (x^{\mu}_l), I_{\cal L} (x^{\mu}_l), I_S (x^{\mu}_l), I_{\psi} (x^{\mu}_l), I_{\phi} (x^{\mu}_l), I_A^{\mu} (x^{\mu}_l), I_x^{\mu} (x^{\mu}_l))  + \nonumber \eeq
\beq + \{1- T_{\epsilon_T}[I_q (x^{\mu}_l, t) \in (q'_l-1/2, q'_l+1/2)] \} \times \nonumber \eeq
\beq \times (q_{2l}, q'_{2l}, {\cal L}_{u2l}, S_{2l}, \psi_{2l}, \phi_{2l}, A^{\mu}_{2l}, x^{\mu}_{2l}) \big\} T_{\epsilon_T} [I_P (x_l^{\mu}) \in (9/2,11/2) ] +  \eeq
\beq +  \big\{-(q_{4l}, q'_{4l}, {\cal L}_{4l}, S_{4l}, \psi_{4l}, \phi_{4l}, A^{\mu}_{4l}, x^{\mu}_{4l}) +  \nonumber \eeq
\beq + T_{\epsilon_T} [I_q (x^{\mu}_l, t) \in (q_l-1/2, q_l+1/2)] \times \nonumber \eeq
\beq \times (I_q (x^{\mu}_l), I_q' (x^{\mu}_l), I_{\cal L} (x^{\mu}_l), I_S (x^{\mu}_l), I_{\psi} (x^{\mu}_l), I_{\phi} (x^{\mu}_l), I_A (x^{\mu}_l),  I_x^{\mu} (x^{\mu}_l)) + \nonumber \eeq
\beq + \{1-T_{\epsilon_T} [I_q (x^{\mu}_l, t) \in (q_l-1/2, q_l+1/2)]\} \times \nonumber \eeq
\beq \times (q_{3l}, q'_{3l}, {\cal L}_{3l}, S_{3l}, \psi_{3l}, \phi_{3l}, A^{\mu}_{3l}, x^{\mu}_{3l}) \big\} T_{\epsilon_T} (I_P (x_l^{\mu})\in [13/2, 15/2])    \nonumber \eeq

\beq \epsilon_{\rightarrow} \frac{d (q_{1l}, q'_{1l}, {\cal L}_{1l}, \psi_{1l}, \phi_{1l}, A^{\mu}_{1l}, x^{\mu}_{1l}) }{d \tau} = (-(q_{1l}, q'_{1l}, {\cal L}_{1l}, \psi_{1l}, \phi_{1l}, A^{\mu}_{1l}, x^{\mu}_{1l})  +  \nonumber \eeq
\beq +  (q_{4l}, q'_{4l}, {\cal L}_{4l}, S_{4l}, \psi_{4l}, \phi_{4l}, A^{\mu}_{4l}, x^{\mu}_{4l})) T_{\epsilon_T} (I_P (x_l^{\mu})  \in [7/2,9/2]) \eeq
\beq \epsilon_{\rightarrow} \frac{d (q_{2l}, q'_{2l}, {\cal L}_{2l}, S_{2l}, \psi_{2l}, \phi_{2l}, A^{\mu}_{2l}, x^{\mu}_{2l})}{d \tau} = (-(q_{2l}, q'_{2l}, {\cal L}_{2l}, S_{2l}, \psi_{2l}, \phi_{2l}, A^{\mu}_{2l}, x^{\mu}_{2l}) +  \nonumber \eeq
\beq + (q_{4l}, q'_{4l}, {\cal L}_{4l}, S_{4l}, \psi_{4l}, \phi_{4l}, A^{\mu}_{4l}, x^{\mu}_{4l})) T_{\epsilon_T} (I_P(x_l^{\mu}) \in [11/2,13/2]) \eeq
\beq \epsilon_{\rightarrow} \frac{d (q_{3l}, q'_{3l}, {\cal L}_{3l}, S_{3l}, \psi_{3l}, \phi_{3l}, A^{\mu}_{3l}, x^{\mu}_{3l})}{d \tau} = (-(q_{3l}, q'_{3l}, {\cal L}_{3l}, S_{3l}, \psi_{3l}, \phi_{3l}, A^{\mu}_{3l}, x^{\mu}_{3l}) +  \nonumber \eeq
\beq + (q_{4l}, q'_{4l}, {\cal L}_{4l}, S_{4l}, \psi_{4l}, \phi_{4l}, A^{\mu}_{4l}, x^{\mu}_{4l})) T_{\epsilon_T} (I_P(x_l^{\mu}) \in [15/2,17/2]) \eeq

\subsection*{9. Conclusion}

The main thing that distinguishes this work from other papers on interpretation of quantum mechanics is that we have shifted the focus away from the "measurement" and towards the definition of wave function as such. As quotes in \cite{PhysicsForums} show, the wave function loses its ontological meaning due to the presence of configuration space. Thus, we have proposed a way of removing configuraiton space altogether; we then invented a "classical" dynamics in "ordinary space" in such a way that the mathematical outcomes of calculations "based on configuration space" ultimately emerge. Our dynamics is "classical" in a sense that there is no instant communication but there is "superluminal" one, where "superluminal" speed is still finite. For that purpose, we have borrowed the idea of bimetric framework suggested by John Moffat, \cite{Bimetric1}, Vitalij Garber, \cite{Bimetric2}, Gisin, \cite{Bimetric3} and possibly others. Furthermore, we have proposed a way in which bimetric framework emerges from "single metric" one where "superluminal" metric is fundamental while "speed of light" as we know it is emergent outcome of the lattice that "floats" in a continuum (similar to speed of sound being emergent in the air).

The fundamental assumption of our work is that superluminal speed, $c_{\rm S}$, is large enough for the signals to cross the entire lattice (whose size is also finite) within a ``very small time". This, however, might not necessarely be the case. If, within said "small" period of time, the superluminal signals pass a large region, such as a galaxy, but not the entire universe, new physics will emerge. Possibly, we will obtain some other version of QFT, which is based on several Hilbert spaces (rather than just one) over various overlapping ``smaller domains". These domains, however, will be of a very large size; thus when we are performing experiments in our laboratory we will observe QFT based on a single domain, leading to conventional physics predictions. 

This logically leads us to ask  whether or not this ``different QFT" will make some cosmological predictions that conflict with current QFT. Incidentally, there has already been experimental attempts to detect the finiteness of superluminal speed on large enough scales (see, for example, \cite{Experiments}). However, while they overfocus on the issue of measurement itself, they neglect the fact that new physics will be operating even without the measurement, for the above stated reasons. In other words, they assume that "configuration space" holds true until collapse and yet the collapse itself progresses with finite speed. From our point of view,  this is logically inconsistent. On the other hand, the work presented in this paper will allow us to make predictions for more consistent scenario, where both configuration space, and measurement non-locality, break down at the same time. This will ultimately allow us to make a better predictions of the "unconventioinal" experimental outcome one is looking for, as opposed to merely falisfying the conventional one.  

Admittedly, it is still the case that some sort of theory of measurement has to be "added" to what we have done. However, in light of the fact that Bohmian and GRW collapse models already exist, they can simply be "converted" into our framework in the similar fashion as we "converted" the measurement-free unitary evolution. This would not have any bearing on reliability of these models (such as the correct estimate of classical scale); rather this would "conceptually" convert them into "classical physics". Thus, whatever questions remain, are of the level of "empirical evidence in favor of classical theory" as opposed to "conceptual difficulties with quantum theory". In this paper we have been exclusively focusing on measurement-free evolution, while the conversion of theories of measurements are left for future work (although one proposal was made in Chapter 5 of \cite{Long}).

\end{document}